\documentclass[intlimits,twoside,a4paper]{article}

\usepackage[cp1251]{inputenc}

\usepackage[eqsecnum]{cmpj3}

\usepackage{bm}

\DeclareMathOperator{\erf}{erf}

\allowdisplaybreaks

\issue{2018}{21}{3}{33707}
\doinumber{10.5488/CMP.21.33707}

\title{Interpreting pulse-shape effects in pump-probe spectroscopies}
\author[A.M. Shvaika, O.P. Matveev, T.P. Devereaux, J.K. Freericks]{A.M. Shvaika\refaddr{ICMP}, O.P. Matveev\refaddr{ICMP}, T.P. Devereaux\refaddr{Geballe,SIMES}, J.K. Freericks\refaddr{GU}}
\addresses{
\addr{ICMP} Institute for Condensed Matter Physics of the National Academy of Sciences of Ukraine,\\ 1 Svientsitskii St., 79011 Lviv, Ukraine
\addr{Geballe} Geballe Laboratory for Advanced Materials, Stanford University, Stanford, CA 94305, USA
\addr{SIMES} Stanford Institute for Materials and Energy Sciences (SIMES), SLAC National Accelerator Laboratory,\\ Menlo Park, CA 94025, USA
\addr{GU} Department of Physics, Georgetown University, 37th and O Streets, NW, Washington, DC 20057, USA
}

\date{Received August 15, 2018}

\begin{document}
	
	\maketitle
	
	\begin{abstract} 
		The effect of the pulse-shape on pump-probe spectroscopies is examined for the simplest model of noninteracting fermions on an infinite-dimensional hypercubic lattice.  The probe-modified density of states follows the time evolution of the pump and displays narrowing and Floquet-like sidebands at the pump maximum, whereas the photoelectron spectra are also strongly affected by the nonequilibrium occupation of the single-particle states due to the excitation from the pump. The nonequilibrium Raman cross section is derived, and the nonresonant one in both the $A_{1\mathrm{g}}$ and $B_{1\mathrm{g}}$ symmetries contains a number of peaks at the pump maximum, which can be attributed to an interference effect or Brillouin scattering off the time variations of the stress tensor. Both the ``measured'' occupation of single-particle states and the ratio of Stokes to anti-Stokes peaks are strongly modified by the probe-pulse width, which must be included in the interpretation of experimental results.
		\pacs 78.47.J-, 79.60.-i, 78.30.-j, 71.10.Fd
	\end{abstract}

\section{Introduction}

Time-resolved spectroscopy is a powerful tool to investigate the dynamical properties of quantum materials at their inherent time-scales~\cite{rossi:895,giannetti:58,hsieh:035128,hellmann:187401,perfetti:067402,wegkamp:216401}. In most cases, it is employed in a pump-probe setup --- first a pump excites the system into a nonequilibrium state and then the probe measures the property of interest. Using pulsed lasers in different spectral regions for the pump, one can select particular excitation modes to be resonantly driven by tuning the driving frequency to the excitation energy. There are many different probes that can be employed depending on whether one is measuring scattered electrons via photoemission spectroscopy (PES) or angle-resolved photoemission spectroscopy (ARPES) or is measuring scattered photons in infrared (IR) spectroscopy, or X-ray absorption spectroscopy (XAS), or many others. We want to highlight one recent study on the time-resolved phononic Raman scattering~\cite{yang:40876}, which was combined with time-resolved ARPES to separately determine electronic and phononic temperatures in graphite. This is an example, which is becoming increasingly common, of an experiment that combines multiple probes on the same material in order to learn more about its nonequilibrium relaxation dynamics.

There is a fair amount of theoretical work on these problems --- we have considered  time-resolved PES~\cite{freericks:98351F,matveev:115167,freericks:034007} and nonresonant electronic Raman scattering~\cite{freericks:1063807} for the Falicov-Kimball model~\cite{falicov:997}, which is the simplest strongly correlated electronic model  that has an exact solution within  dynamical mean-field theory (DMFT)~\cite{freericks:1333}. Obviously, whenever one performs a pump-probe experiment, there arises a question: to what extent does the shape of the pump or probe pulse affect the results of the
experiment, and how can we best compensate for these effects if we want to understand the behaviour of the system unmodified by artifacts of the experimental measurement process? The most obvious issue
arises from the width of the pulses, because frequency and time are related via a Fourier transformation and hence they obey energy-time uncertainty relations~\cite{randi:115132}. However, there are also effects due to the pulse amplitude (which can even create inverted populations~\cite{golez:246402}) and the shape of the envelope. A related question is: how do the probes affect the ability to extract effective temperatures for the excited system? A number of experiments have employed different methods to determine these effective temperatures~\cite{yang:40876,smallwood:235107,stafford:245403,rohde2018decoding}.

In order to determine the effect of the pulse shape on experimental results, we consider the case of  noninteracting fermions on a $D$-dimensional hypercubic lattice. We take the DMFT limit of $D\to\infty$, which, on the one hand, allows for a comparison with the previous DMFT results and, on the other hand, makes it sometimes possible to obtain analytic results. The organization of the paper is as follows: In section~\ref{sec:Hamiltonian}, we present our model. Section~\ref{sec:1particle} considers the single-particle properties, i.e.,~the density of states (DOS) and the time-resolved PES signal, while section~\ref{sec:2particle} considers nonresonant electronic Raman scattering, which measures two-particle (collective bosonic) excitations. We conclude in section~\ref{sec:sec:conclusions}.
	
\section{Hamiltonian}\label{sec:Hamiltonian}

We consider noninteracting spinless fermions on a $D$-dimensional hypercubic lattice. The interaction with an electromagnetic field is included through the Peierls substitution~\cite{peierls:763,peierls:763eng,turkowski:085104}
\begin{equation}
\mathcal{H}(t) =- \sum_{ij} t_{ij} \re^{-\ri \int_{\mathbf{R}_j}^{\mathbf{R}_i} \rd\mathbf{r}' \cdot \mathbf{A}(\mathbf{r}',t)} c_i^{\dag} c_j^{\phantom\dagger} .
\end{equation}
The hopping is between nearest neighbours only with a hopping integral given by $t=t^*/2\sqrt{D}$, and $t^*$ is used as the energy unit.
The pump is described by a homogeneous electric field directed along the unit cell diagonal of a $D$-dimensional lattice $\mathbf{E}(t) = (E(t),E(t),E(t),\ldots)$, where
\begin{equation}
E(t)=E_0\cos{(\omega_{\mathrm{p}} t)} \,\re^{-\frac{t^2}{\sigma_{\mathrm{p}}^2}}.
\end{equation}
Here, $\omega_{\mathrm{p}}$ is the pump pulse frequency and $\sigma_{\mathrm{p}}$ is the pump probe width; the pump is always centered at $t=0$.
The total vector potential $\mathbf{A}(t) = (A(t),A(t),A(t),\ldots)$ contains two contributions --- one from the pump pulse and the other one from the probe pulse,
\begin{equation}
A(t)=A_{\text{pump}}(t)+A_{\text{probe}}(t), \qquad A_{\text{pump}}(t) = -\int_{-\infty}^t E(t') \rd t'.
\end{equation}
We describe the probe pulse later; it is small, so it will be treated in perturbation theory via the Kubo response methodology.

We can also write the Hamiltonian in momentum space via
\begin{equation}\label{eq:Hnint}
\mathcal{H}(t) = \sum_{\mathbf{k}} \varepsilon(\mathbf{k}-\mathbf{A}(t)) c_{\mathbf{k}}^{\dag} c_{\mathbf{k}}^{\phantom\dagger}\,,\qquad c_\textbf{k}^{\phantom\dagger}=\frac{1}{\sqrt{N}}\sum_{j=1}^Nc_j^{\phantom\dagger} \re^{\ri \textbf{k}\cdot \textbf{R}_j},
\end{equation}
where $\varepsilon(\mathbf{k}-\mathbf{A}(t))=-\sum_j t_{ij} \exp\{\ri [\textbf{k}- \mathbf{A}(t)]\cdot\mathbf{R}_{ij}\}$ is the band energy. The sum is over all neighbours $j$ of site $i$ and $\textbf{R}_{ij}=\textbf{R}_{i}-\textbf{R}_{j}$. Note that this form of the Hamiltonian allows us to immediately see that the Hamiltonian commutes with itself at different times $[\mathcal{H}(t),\mathcal{H}(t')]=0$. This result makes determining many time-dependent quantities much easier than for systems where the Hamiltonian does not commute with itself at two different times.

\section{DOS, PES, and occupation of the single-particle states}\label{sec:1particle}

Now, we proceed to the calculations of the DOS and the time-resolved PES measured in a pump-probe experiment. Such quantities are defined through the single-particle Green's function and it is convenient to apply the Kadanoff-Baym-Keldysh formalism~\cite{kadanoff:1962,keldysh:1018} in this case. 
The single-particle Green's function on the Schwinger-Keldysh contour is defined by
\begin{equation}
G_{\mathbf{k}}^{\text c}(t,t')=-\ri\left\langle\mathcal{T}_\text{c} c_{\mathbf{k}}^{\phantom\dagger}(t) c_{\mathbf{k}}^{\dag}(t')\right\rangle
\end{equation}
and in the case of noninteracting electrons~\eqref{eq:Hnint}, it takes the form~\cite{turkowski:085104,freericks:266408}
\begin{equation}\label{eq:GFk}
G_{\mathbf{k}}^{\text c}(t,t')=\ri\left[f(\varepsilon(\mathbf{k})-\mu)-\Theta_\text{c}(t,t')\right]  \exp\left[-\ri\int_{t'}^{t} \rd\bar{t}\; \varepsilon(\mathbf{k}-\mathbf{A}(\bar{t})) \right],
\end{equation}
where
\begin{equation}
f(\omega-\mu) = \frac{1}{\re^{\,\beta(\omega-\mu)}+1}
\end{equation} 
is the Fermi-Dirac distribution function, which arises from the initial equilibrium occupation of the single-particle states before the pump, and $\Theta_\text{c}(t,t')$ is the Heaviside step function on the Schwinger-Keldysh contour which equals $1$ when $t$ is ahead of $t'$, $0$ when $t$ is behind of $t'$, and $1/2$ when $t$ coincides with $t'$ on the contour.

For the $D$-dimensional hypercubic lattice with a nearest-neighbour hopping, we have
\begin{align}
\varepsilon(\mathbf{k}-\mathbf{A}(t)) =  \lim_{D\to\infty}\left \{-\frac{t^*}{\sqrt{D}} \sum_{\alpha=1}^{D} \cos[k_{\alpha}-A(t)]\right \}
= \varepsilon(\mathbf{k}) \cos A(t) + \bar{\varepsilon}(\mathbf{k}) \sin A(t),
\end{align}
where
\begin{equation}\label{eq:eedef}
\varepsilon(\mathbf{k}) = \lim_{D\to\infty}\left (-\frac{t^*}{\sqrt{D}} \sum_{\alpha=1}^{D} \cos k_{\alpha}\right ), \qquad \bar{\varepsilon}(\mathbf{k}) = \lim_{D\to\infty}\left (-\frac{t^*}{\sqrt{D}} \sum_{\alpha=1}^{D} \sin k_{\alpha}\right ).
\end{equation}

Since the Green's function depends on the momentum only through the band energy $\varepsilon$ and the projection of the electron velocity onto the electric field direction $\bar\varepsilon$,  we will replace a summation over wavevector by an integration over a joint DOS
\begin{equation}
\frac{1}{N} \sum_{\mathbf{k}} \longrightarrow \int \rd\varepsilon \int \rd\bar{\varepsilon}\; \rho(\varepsilon,\bar{\varepsilon}).
\end{equation}
Furthermore, the momentum-dependent Green's function can be written as 
\begin{equation}\label{eq:Gc}
G_{\varepsilon,\bar{\varepsilon}}^\text{c}(t,t')=\ri\left[f(\varepsilon-\mu)-\Theta_\text{c}(t,t')\right]  \exp\left\{-\ri\int_{t'}^{t} \rd\bar{t}\; [\varepsilon \cos A(\bar{t}) + \bar{\varepsilon} \sin A(\bar{t})] \right\}.
\end{equation}
While the expressions in \eqref{eq:GFk}--\eqref{eq:eedef} are exact for any lattice with  nearest-neighbour hopping, we consider only the DMFT limit $D\to\infty$, with $t^*=1$. In this case, the joint DOS becomes Gaussian~\cite{turkowski:085104}
\begin{equation}
\rho(\varepsilon,\bar{\varepsilon}) = \frac{1}{N} \sum_{\mathbf{k}} \delta(\varepsilon-\varepsilon(\mathbf{k})) \;\delta(\bar{\varepsilon}-\bar{\varepsilon}(\mathbf{k})) = \dfrac{\re^{-\varepsilon^2}}{\sqrt{\piup}} \;\dfrac{\re^{-\bar{\varepsilon}^2}}{\sqrt{\piup}}\,,
\end{equation}
which simplifies calculations and allows us to obtain many analytic results.

Now, with all these preliminaries done, we are ready to consider the effect of the probe pulse width on the single-particle quantities.

\subsection{Equilibrium case}

In equilibrium, with no pump pulse ($A_{\text{pump}}(t)=0$), the retarded Green's function is given by
\begin{equation}
G_{\mathbf{k}}^\text{r}(t-t') =-\ri \Theta(t-t') \exp\left[-\ri \varepsilon(\mathbf{k}) (t-t') \right]
\end{equation}
and if we employ a monochromatic probe beam of infinite width, the Fourier transform of the Green's function is
\begin{equation}
G_{\mathbf{k}}^\text{r}(\omega) =\frac{1}{\omega-\varepsilon(\mathbf{k})+\ri0^+}.
\end{equation}
This yields the local DOS via
\begin{equation}\label{eq:Ad_eq}
A_\text{d}(\omega) = \frac{1}{N} \sum_{\mathbf{k}} \delta(\omega-\varepsilon(\mathbf{k})) = \dfrac{\re^{-\omega^2}}{\sqrt{\piup}}.
\end{equation}
In the same situation, the lesser Green's function is given by
\begin{equation}
G_{\mathbf{k}}^{<}(t-t')=\ri f(\varepsilon(\mathbf{k})-\mu) \exp\left[-\ri \varepsilon(\mathbf{k}) (t-t') \right],
\end{equation}
which Fourier transforms to
\begin{equation}\label{eq:PESeq}
P(\omega) = f(\omega-\mu) A_\text{d}(\omega).
\end{equation}
This result, which is the product of the distribution function times the local DOS, also yields the PES signal (if we neglect matrix-element effects).

In a pump/probe experiment with probe pulses, we always have to make tradeoffs. The pulses should be narrow enough to achieve good temporal selectivity but not too narrow, otherwise they lose all spectral features in frequency space. This is called energy-time uncertainty~\cite{randi:115132}. We analyze this behaviour now.  If we express the probe-pulse vector potential as
\begin{equation}
A_0 \re^{\ri\omega t}s(t;t_0)
\end{equation}
with $A_0$ being the probe vector potential amplitude,  and $s(t;t_0)$ --- the probe envelope function; we assume $A_0$ is small, and it will not enter any of the perturbative results we calculate below. For Gaussian probe pulses, with an envelope function centered at $t_0$ and with width $\sigma_{\mathrm{b}}$, we have
\begin{equation}
s(t;t_0)=\frac{1}{\sigma_{\text{b}}\sqrt{\piup}} \re^{-\frac{(t-t_0)^2}{\sigma^{2}_{\text{b}}}}.
\end{equation}
The Fourier transformation from  time to frequency is then modified by additional factors of $s(t;t_0)$ and $s(t';t_0)$. For example, the DOS will be changed to a probe-modified DOS which equals
\begin{align}
A_\text{d}(\omega;\sigma_{\text{b}}) &= \Im \int \rd t \int \rd t'\; s(t;t_0) s(t';t_0) \re^{\ri\omega(t-t')} \frac{1}{N} \sum_{\mathbf{k}} G_{\mathbf{k}}^\text{r}(t-t')
\nonumber \\
&= \int \rd t \int \rd t'\; s(t;0) s(t';0) \re^{\ri\omega(t-t')} \int \rd\varepsilon \dfrac{\re^{-\varepsilon^2}}{\sqrt{\piup}} \re^{-\ri \varepsilon (t-t') }
\nonumber \\
&= \dfrac{1}{\sqrt{\frac{\sigma_{\mathrm{b}}^2}{2}+1}} \exp\left(-\frac{\omega^2}{1+\frac{2}{\sigma_{\mathrm{b}}^2}}\right).
\label{eq:Ad_sb}
\end{align}
Note that the final result is independent of $t_0$. This is because the rest of the integrand is just a function of $t-t'$, so one can remove $t_0$ by the following shifts in the integration variables: $t\to t+t_0$ and $t'\to t'+t_0$.
Due to the final width of the probe pulses, this probe-modified DOS is not normalized. One can see that the initial Gaussian DOS \eqref{eq:Ad_eq} remains Gaussian, but with a wider bandwidth given by $\sqrt{1+\frac{2}{\sigma_{\mathrm{b}}^2}}$ instead of 1. Note that this pulse-modified DOS is not easily measured, so we examine more experimentally relevant quantities below.

Similarly, when we calculate the spectral density or probe-modified PES, we obtain~\cite{freericks:136401}
\begin{align}
P(\omega;\sigma_{\text{b}}) &= \int \rd t \int \rd t'\; s(t;t_0) s(t';t_0)\, \re^{\ri\omega(t-t')} \frac{1}{N} \sum_{\mathbf{k}} G_{\mathbf{k}}^{<}(t-t')
\nonumber \\
&= \int \rd\varepsilon f(\varepsilon-\mu) \dfrac{\re^{-\varepsilon^2}}{\sqrt{\piup}} \re^{-\sigma_{\mathrm{b}}^2(\omega-\varepsilon)^2/2}.
\end{align}
For a monochromatic probe beam ($\sigma_{\mathrm{b}}\to\infty$), the Gaussian factor that depends on $\omega$ becomes a $\delta$-function and one recovers the result in equation~\eqref{eq:PESeq}. On the other hand, for narrow probe pulses $\sigma_{\mathrm{b}}\to0$, the spectral density or PES becomes its average value,  averaged over all frequencies. Since the Fermi-Dirac distribution function approaches a unit step function at zero temperature $(\beta\to\infty$),  the integral can be evaluated and we find an expression similar to equation~\eqref{eq:PESeq} for the pulse-modified PES
\begin{equation}\label{eq:PESeqprobe}
P(\omega;\sigma_{\text{b}}) = f(\omega,\mu; \sigma_{\text{b}}) A_\text{d}(\omega;\sigma_{\text{b}}),
\end{equation}
where $A_\text{d}(\omega; \sigma_{\text{b}})$ is defined by \eqref{eq:Ad_sb} and instead of the step-like Fermi-Dirac distribution function, we find a smoothed function given by
\begin{equation}\label{eq:FDprobe}
\vspace{-1mm}
f(\omega,\mu;\sigma_{\text{b}}) = \frac12 + \frac12 \erf\left\{\frac{\sigma_{\text{b}}^2}{\sqrt{4+2\sigma_{\text{b}}^2}} \left[\left(1+\frac{2}{\sigma_{\text{b}}^2}\right)\mu-\omega\right]\right\}.
\end{equation}
At zero temperature, the probe-modified PES has a smoothed step-like feature that is located at a shifted Fermi level! Such a smoothed function looks similar to the Fermi-Dirac distribution but at a higher temperature.  From the slope of $f(\omega,\mu;\sigma_{\text{b}})$ at the Fermi level $(1+{2}/{\sigma_{\text{b}}^2})\mu$, we can estimate this effective inverse temperature (introduced from the finite width of the probe pulse) via $\beta = {4\sigma_{\text{b}}^2}/\Bigl({\sqrt{\piup}\sqrt{4+2\sigma_{\text{b}}^2}}\,\Bigr)$. This implies that the presence of a probe pulse mimics the behaviour of thermal excitations when we examine a PES signal at $T=0$. 

We can also evaluate the probe-modified PES in the limit of infinite temperature $(\beta\to\ 0)$. Here, the Fermi-Dirac distribution function is replaced by a constant equal to the filling, because every state is occupied with the same probability. Then, the PES signal is simply equal to the filling multiplied by the probe-modified DOS. In this case, while the probe affects the DOS, it has no effect on the distribution function. From these results, we conjecture that the effect of the probe on the distribution function is the greatest at low temperatures and disappears as we go to higher temperatures.

Surely these behaviours will play a role when we try to extract effective temperatures for nonequilibrium cases too. We examine this point next.

\subsection{Nonequilibrium case}

In the nonequilibrium case, the Peierls' substitution shifts the momentum label of the different energies in the bandstructure as a function of time. Note that the complete set of energy eigenvalues is unchanged, meaning if we diagonalized the instantaneous Hamiltonian at any given moment, the set of eigenvalues would not change. But the labels do. Since the degeneracy structure of the bands in momentum space influences the DOS, we expect the DOS to change as a function of time due to this relabelling. However, since the DOS is defined via the Green's function, the DOS is also affected by the time dependence of the wavefunctions, which enter the equation of motion for the Green's functions. This means that we should not interpret the transient nonequilibrium DOS as simply being the DOS of the instantaneous Hamiltonian. It is a more complex object. In addition, the driving fields can change the distribution of electrons amongst these states. Fortunately, all of these effects can be treated exactly in a noninteracting system.

In the presence of a pump, the time-dependent DOS is normally defined through the retarded Green's function. When we make a Fourier transform to express it as a function of frequency and some time (due to the fact that it changes with time), we have a number of different choices we can make. One of the most common choices is to perform a Wigner transformation to average and relative times and perform the Fourier transform with respect to relative time. This has the advantage of a clearly defined ``time'' at which we have constructed the DOS, but the noncausal nature of this time can make the interpretation of this DOS confusing. For example, even for an average time before the pump, we will have some relative times where one of the original times $t$ or $t'$ will be after the pump. If these times contribute significantly to the Fourier transform, one will see an effect of the pump on the DOS at an average time before the pump is applied. In addition, there is no guarantee that the DOS defined in this fashion is nonnegative. There are alternatives one can make as well. One can instead fix $t'$ and set $t=t'+\Delta t$ and perform the Fourier transform with respect to $\Delta t$. This produces a causal structure to the system, since only times after $t'$ are involved in the result. However, this DOS is not guaranteed to be nonnegative either. It is also unclear
what time we should associate with this DOS. Here, we instead define a probe-modified DOS in analogy to the probe-modified DOS and PES we worked with above. It is
\begin{equation}
A_\text{d}(\omega;t_0) = \int \rd t \int \rd t'\; s(t;t_0) s(t';t_0)\, \re^{\ri\omega(t-t')} \re^{-\frac14\phi^2(t,t')}  \re^{-\frac14\psi^2(t,t')},
\end{equation}
where the pump field enters through the quantities
\begin{equation}\label{eq:phipsi}
\phi(t,t') = \int_{t'}^{t} \rd\bar{t} \cos A(\bar{t})\quad\text{and}\quad \psi(t,t') = \int_{t'}^{t} \rd\bar{t} \sin A(\bar{t}).
\end{equation}
One can see that the probe-modified DOS does not depend on temperature (as is expected for  noninteracting fermions). More importantly, one can immediately verify that the probe-modified DOS is 
nonnegative. It is also clearly associated with the time $t_0$, although this is still a bit fuzzy due to the finite probe widths and the energy-time uncertainty relations.
Similarly, the spectral density or PES signal is equal to~\cite{freericks:136401}
\begin{align}
P(\omega;t_0) &= \int \rd t \int \rd t'\; s(t;t_0) s(t';t_0)\, \re^{\ri\omega(t-t')} \int \rd\varepsilon \int \rd\bar{\varepsilon}\; \rho(\varepsilon,\bar{\varepsilon})\; G_{\varepsilon,\bar{\varepsilon}}^{<}(t,t')
\nonumber\\
&= \int \rd t \int \rd t'\; s(t;t_0) s(t';t_0)\, \re^{\ri\omega(t-t')} \re^{-\frac14\psi^2(t,t')} \int \rd\varepsilon \;\dfrac{\re^{-\varepsilon^2}}{\sqrt{\piup}} f(\varepsilon-\mu) \re^{-\ri\varepsilon\phi(t,t')},
\end{align}
which is also manifestly nonnegative.
The time delay between the pump and probe pulses is set by $t_0$, since the pump is centered at the origin in time.

In equilibrium, we employed the ratio of the PES to the DOS to determine the distribution function in equation~\eqref{eq:PESeq}. We generalized this result to take into account the probe in equation~\eqref{eq:PESeqprobe}. Motivated by these results, we define the probe-modified nonequilibrium occupation of single-particle states to be the ratio of the probe-modified nonequilibrium PES to the probe-modified nonequilibrium DOS~\cite{smallwood:235107,stafford:245403}
\begin{equation}
n_\text{d}(\omega;t_0) = \dfrac{P(\omega;t_0)}{A_\text{d}(\omega;t_0)}.
\end{equation}
One can estimate an effective temperature at $t_0$  either from the slope of $n_\text{d}(\omega,t_0)$ at the chemical potential or from a least squares (LSQ) interpolation of $n_\text{d}(\omega;t_0)$ with the Fermi-Dirac distribution function (there are other definitions one could use for the effective temperature, but these two are the simplest ones to use, and we examine them thoroughly in this paper).

\begin{figure*}[!t]
	\centering
	\includegraphics[width=0.4\linewidth]{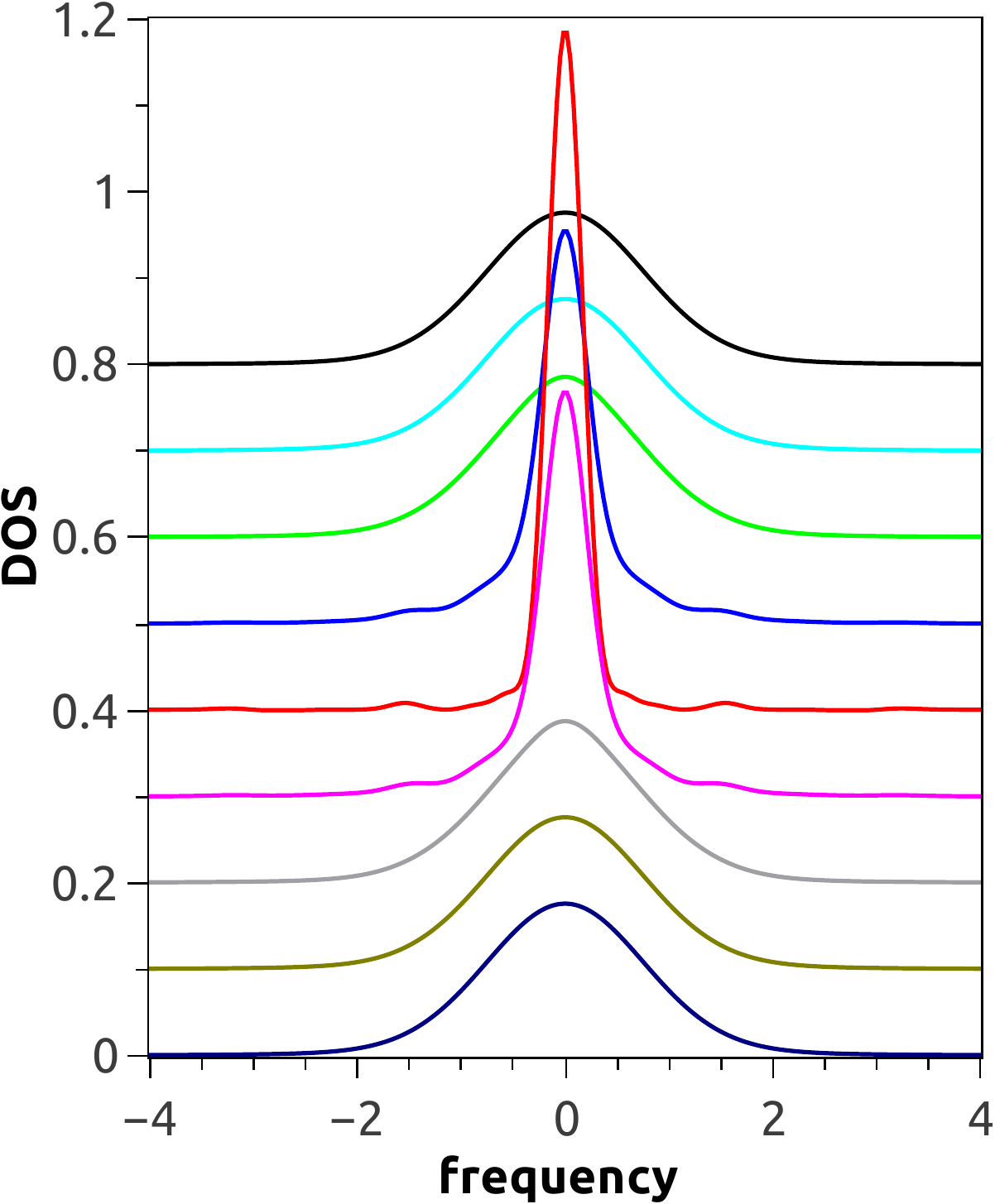}\hfill\includegraphics[width=0.4\linewidth]{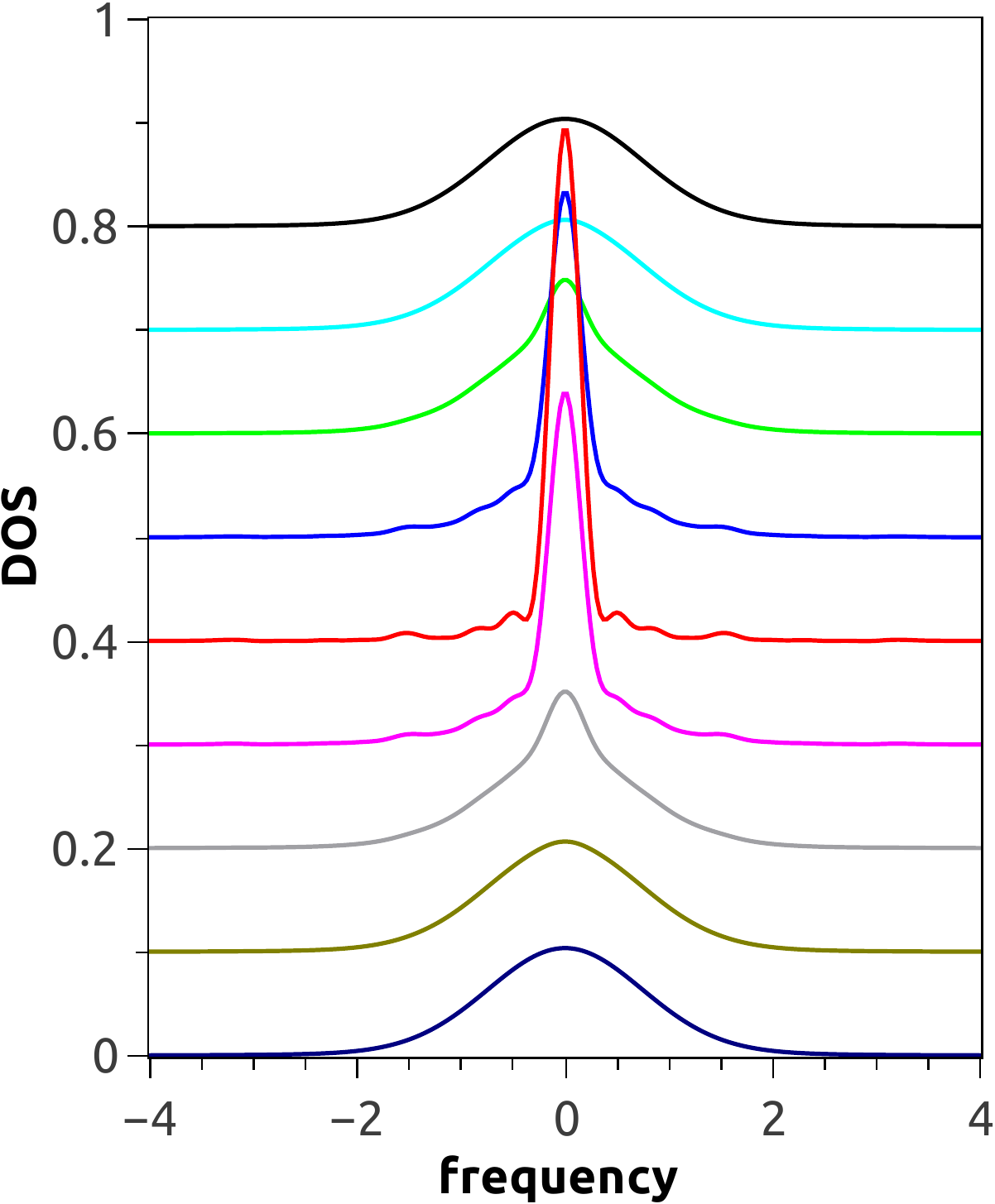} \\ [1em]
	\includegraphics[width=0.4\linewidth]{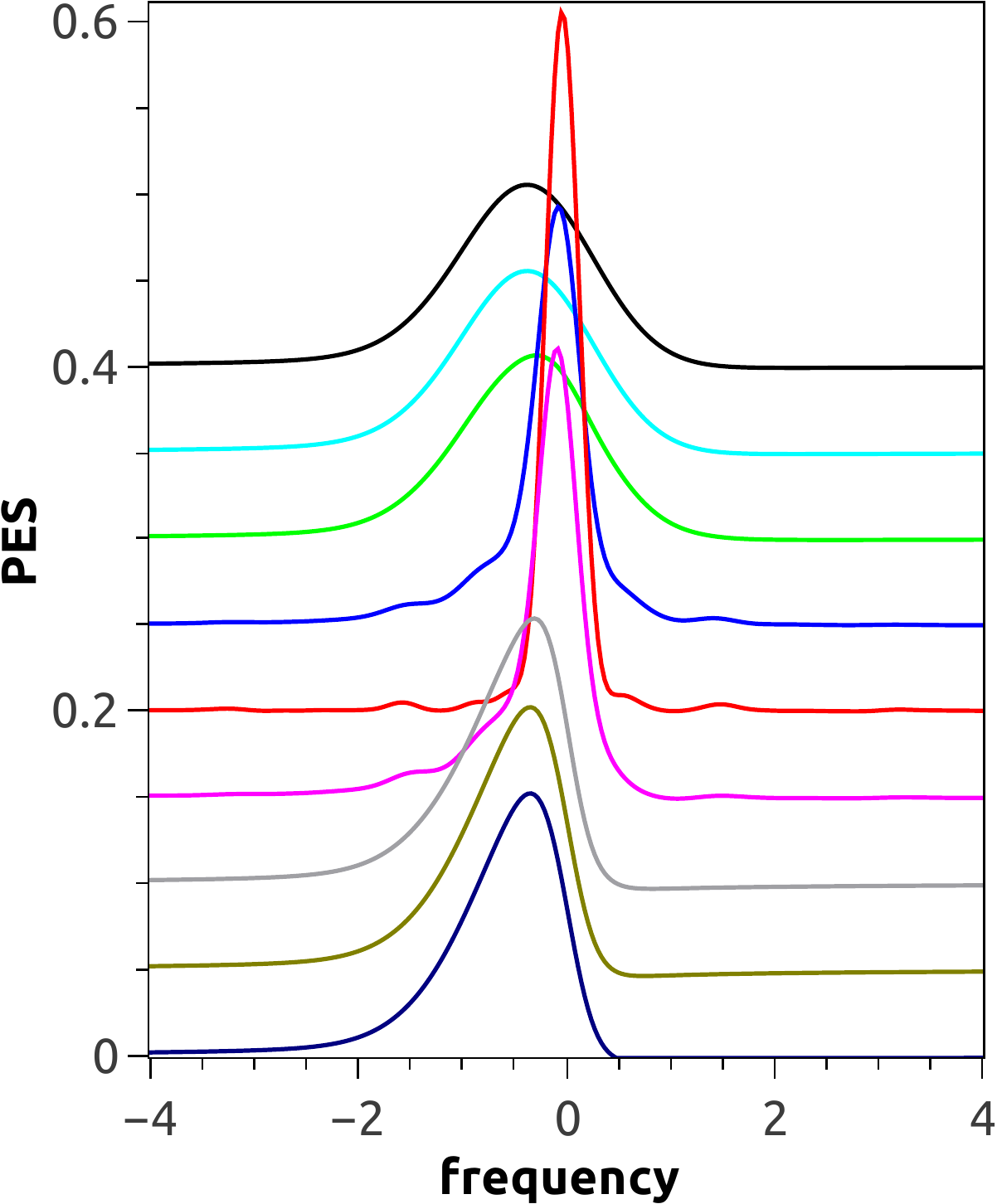}\hfill\includegraphics[width=0.4\linewidth]{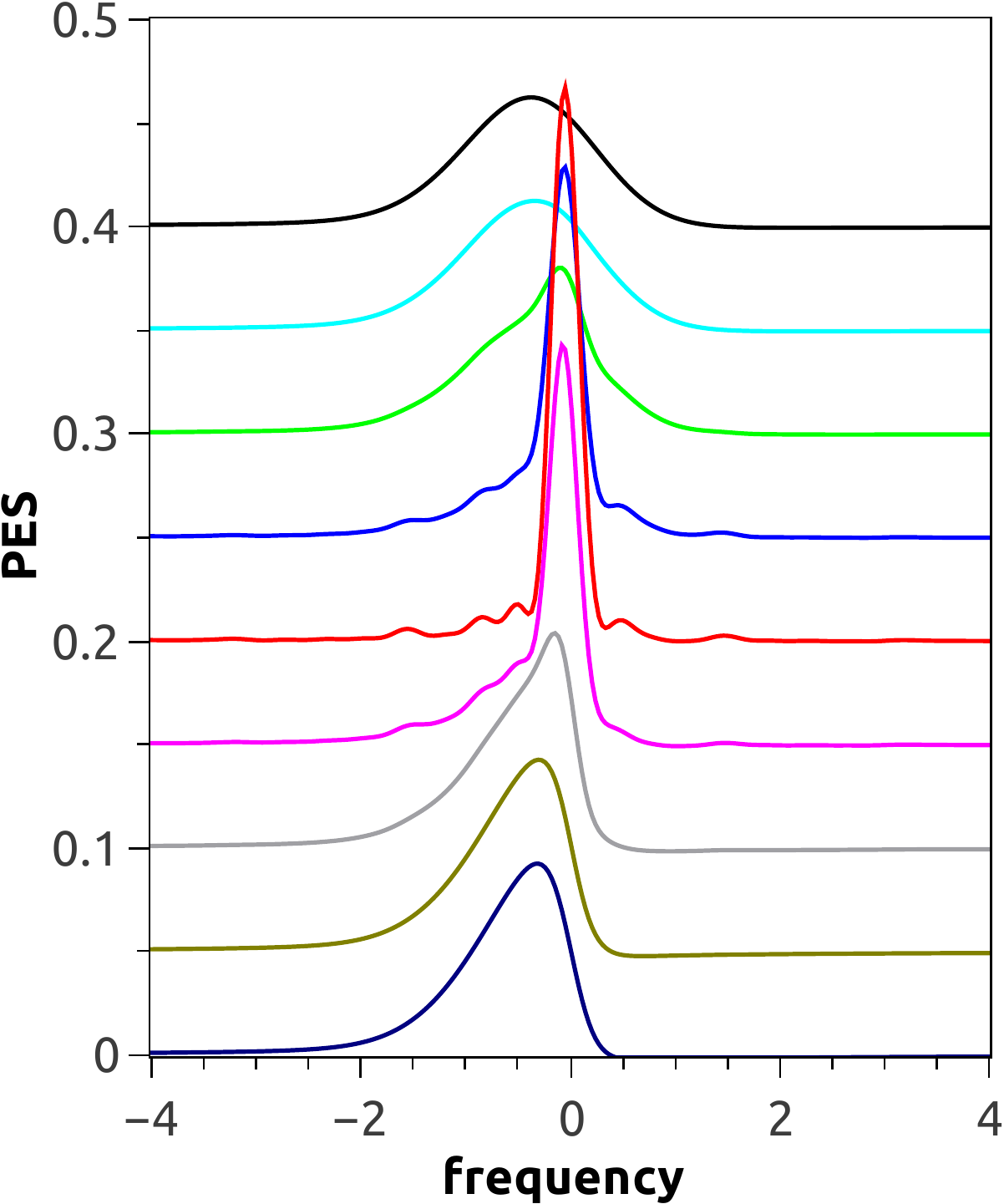}  \\ [2em]
	\includegraphics[width=0.49\linewidth]{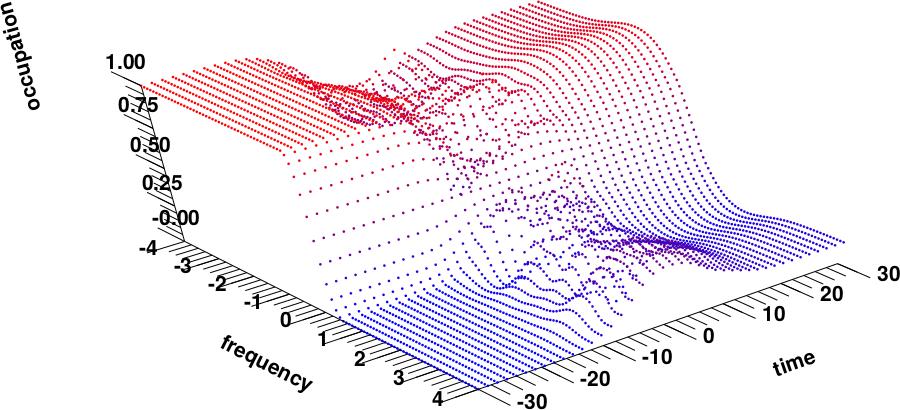}\hfill\includegraphics[width=0.49\linewidth]{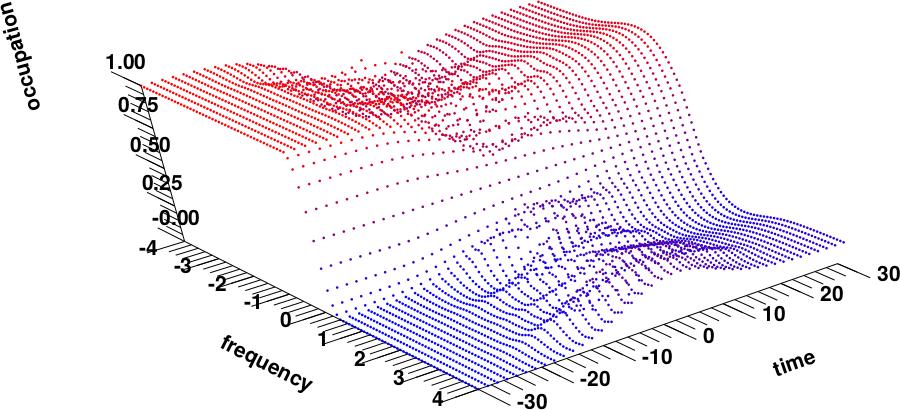} 
	\bigskip
	\caption{(Colour online) Waterfall images of the time-resolved DOS and PES data, plotted for different delay times $t_0\in[-30,30]$ and offset for clarity, and occupation of single-particle states $n_\text{d}(\omega;t_0)$ for $E_0=30$ and $\sigma_{\text{b}}=7$ (left) and $\sigma_{\text{b}}=12$ (right).}
	\label{fig:pese030}
\end{figure*}

\begin{figure*}[!t]
	\centering
	\includegraphics[width=0.4\linewidth]{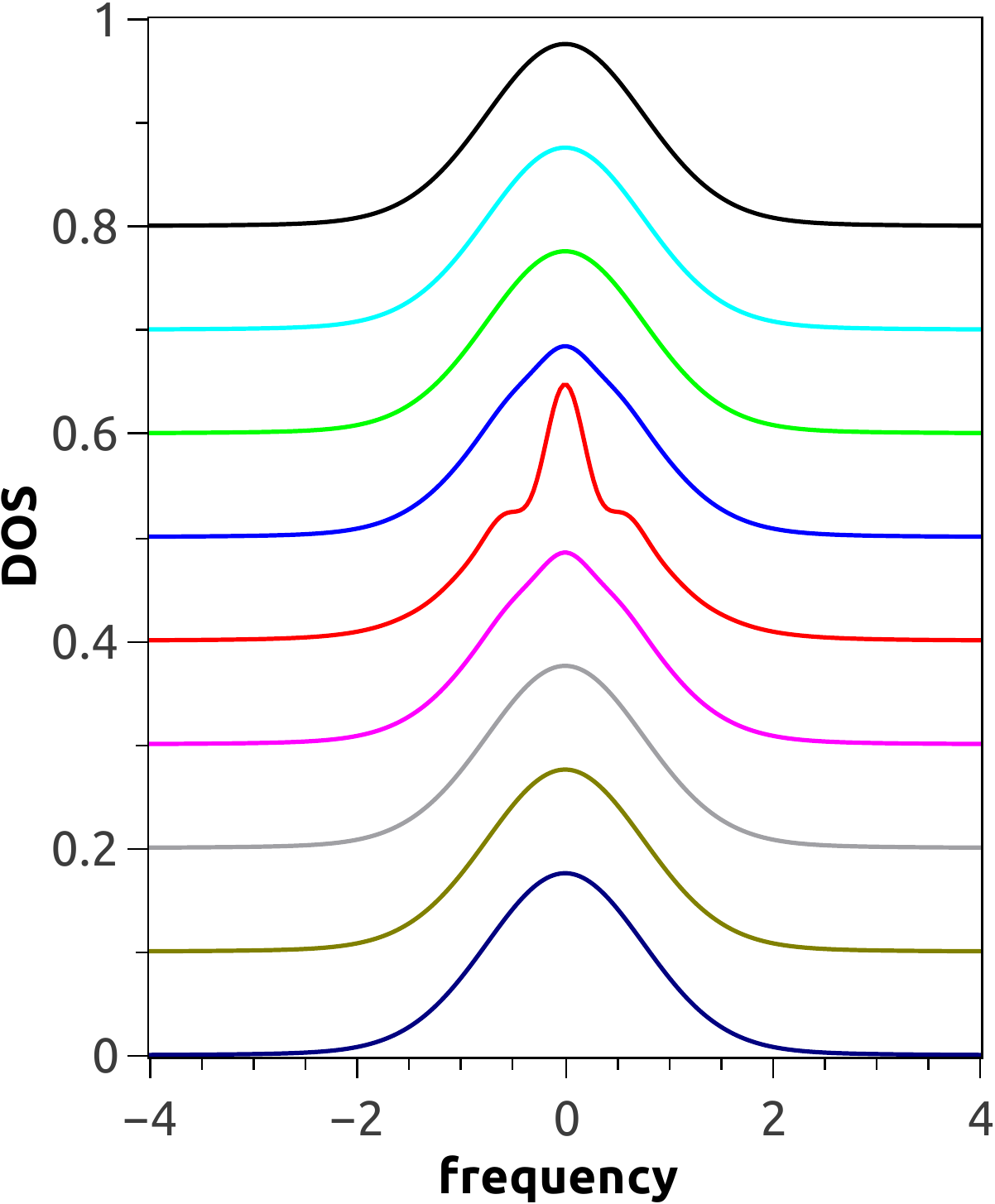}\hfill\includegraphics[width=0.4\linewidth]{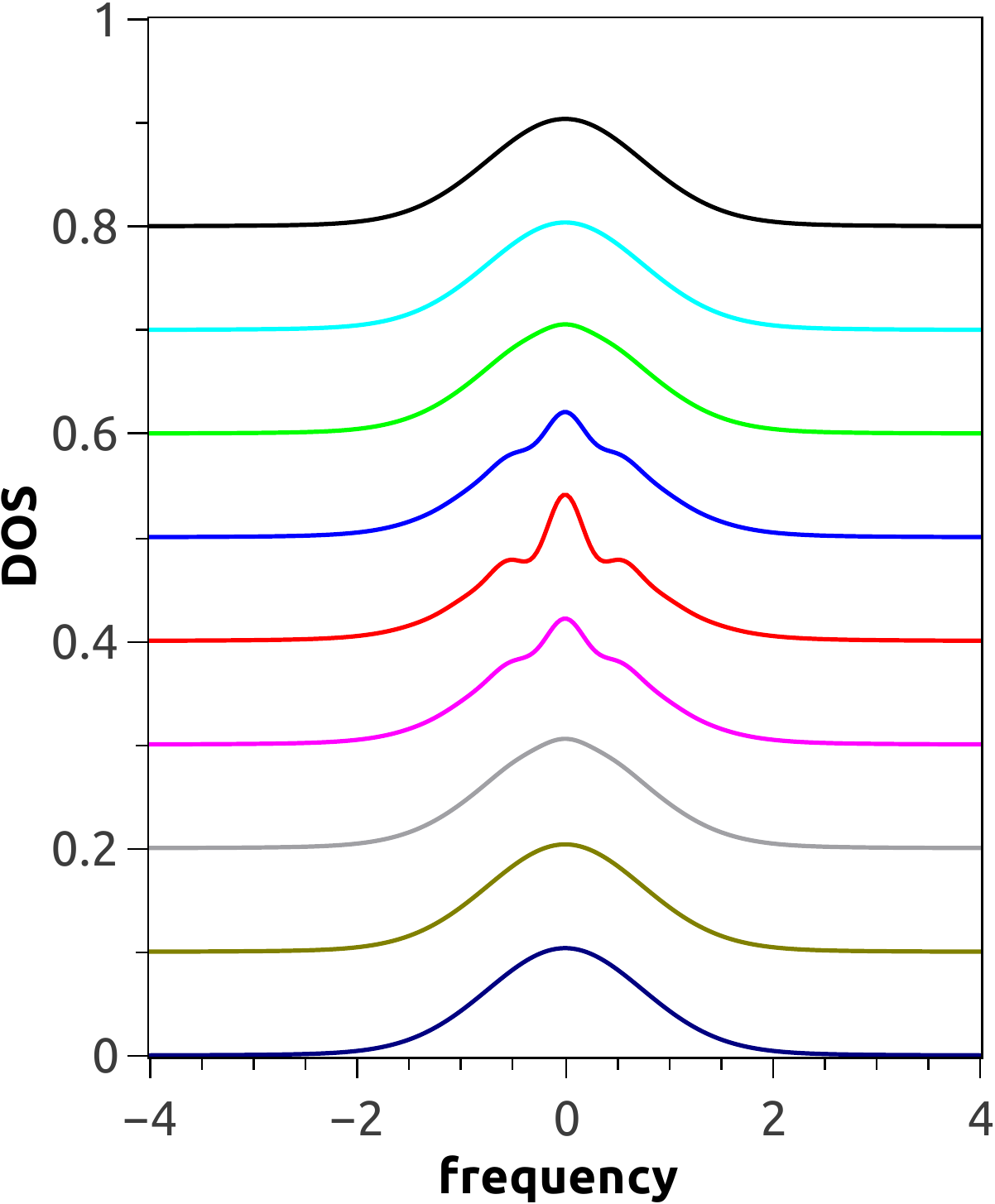} \\ [1em]
	\includegraphics[width=0.4\linewidth]{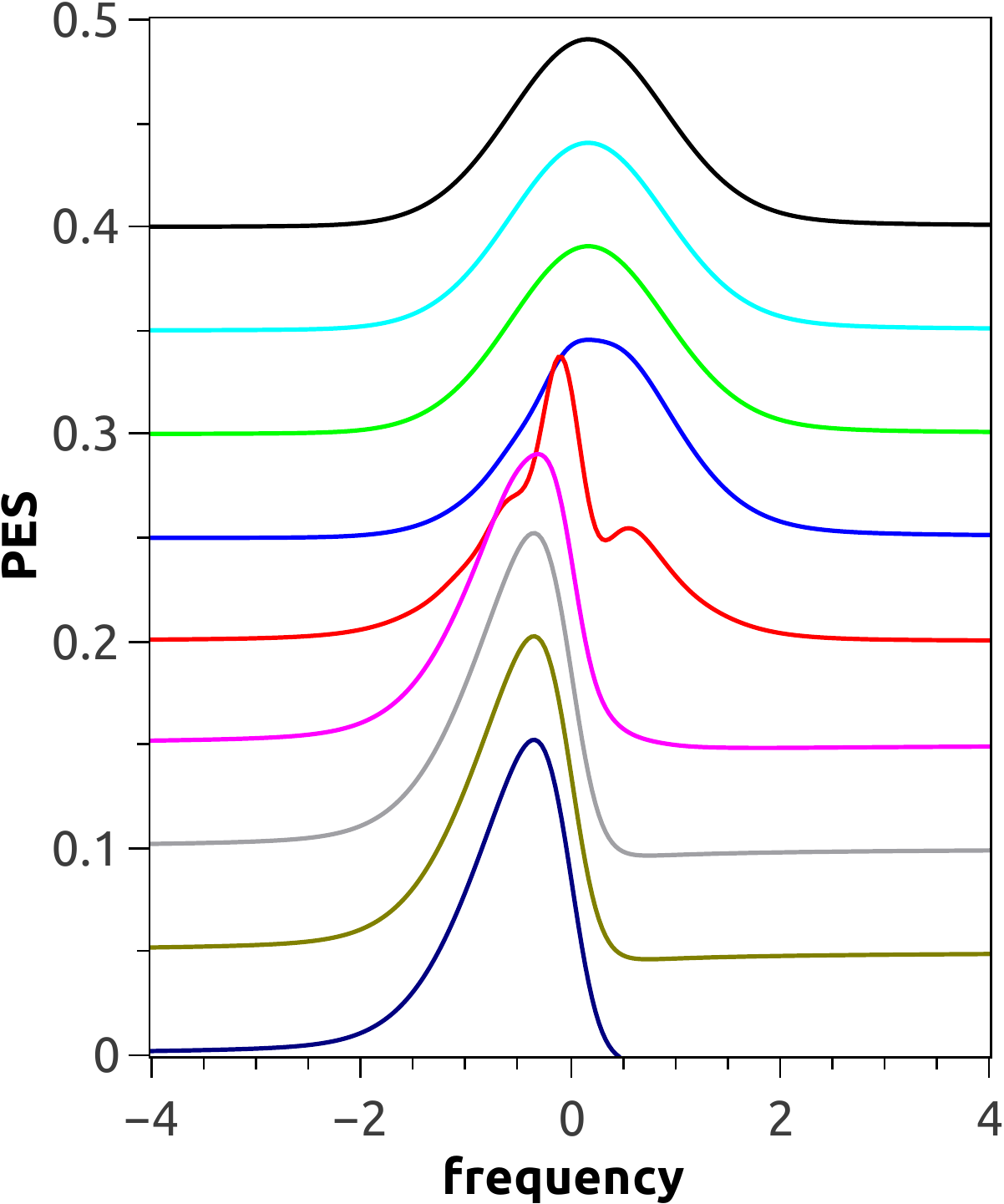}\hfill\includegraphics[width=0.4\linewidth]{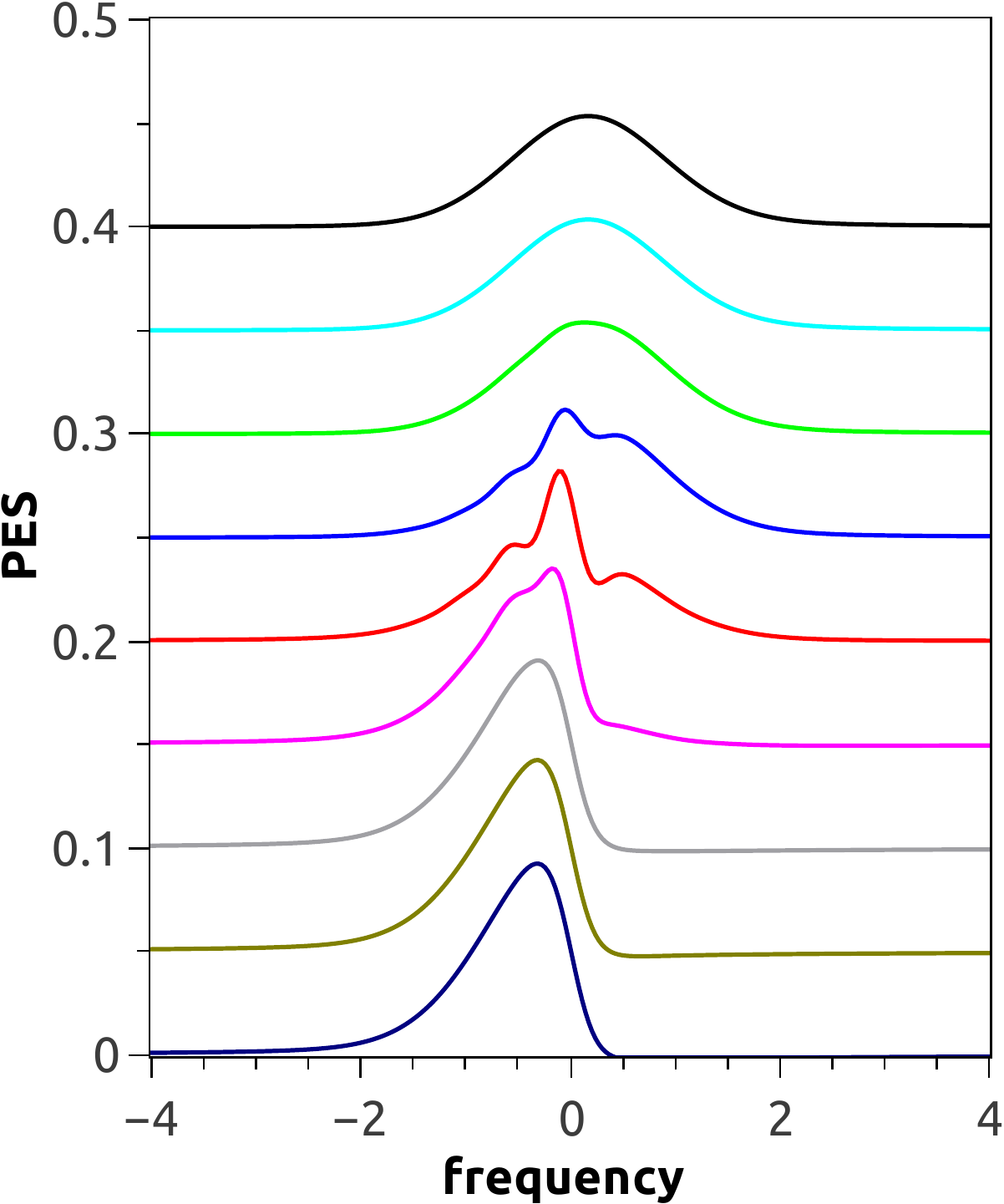} \\ [2em]
	\includegraphics[width=0.49\linewidth]{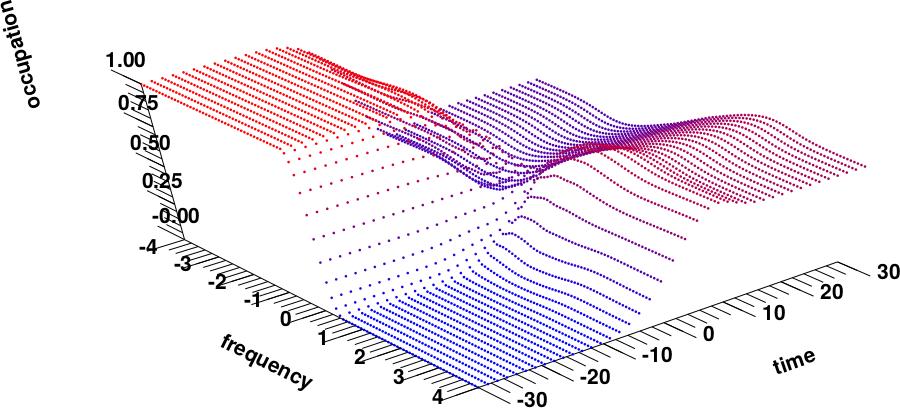}\hfill\includegraphics[width=0.49\linewidth]{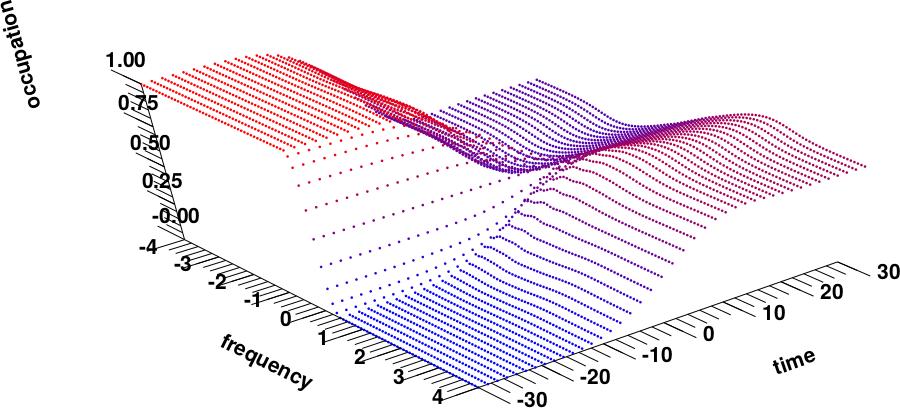} 
	\caption{(Colour online) Waterfall images of the time-resolved DOS and PES data, plotted for different delay times $t_0\in[-30,30]$ and offset for clarity, and occupation of single-particle states $n_\text{d}(\omega;t_0)$ for $E_0=1$ and $\sigma_{\text{b}}=7$ (left) and $\sigma_{\text{b}}=12$ (right).}
	\label{fig:pese001}
\end{figure*}

\begin{figure*}[!t]
	\centering
	\includegraphics[width=0.49\linewidth]{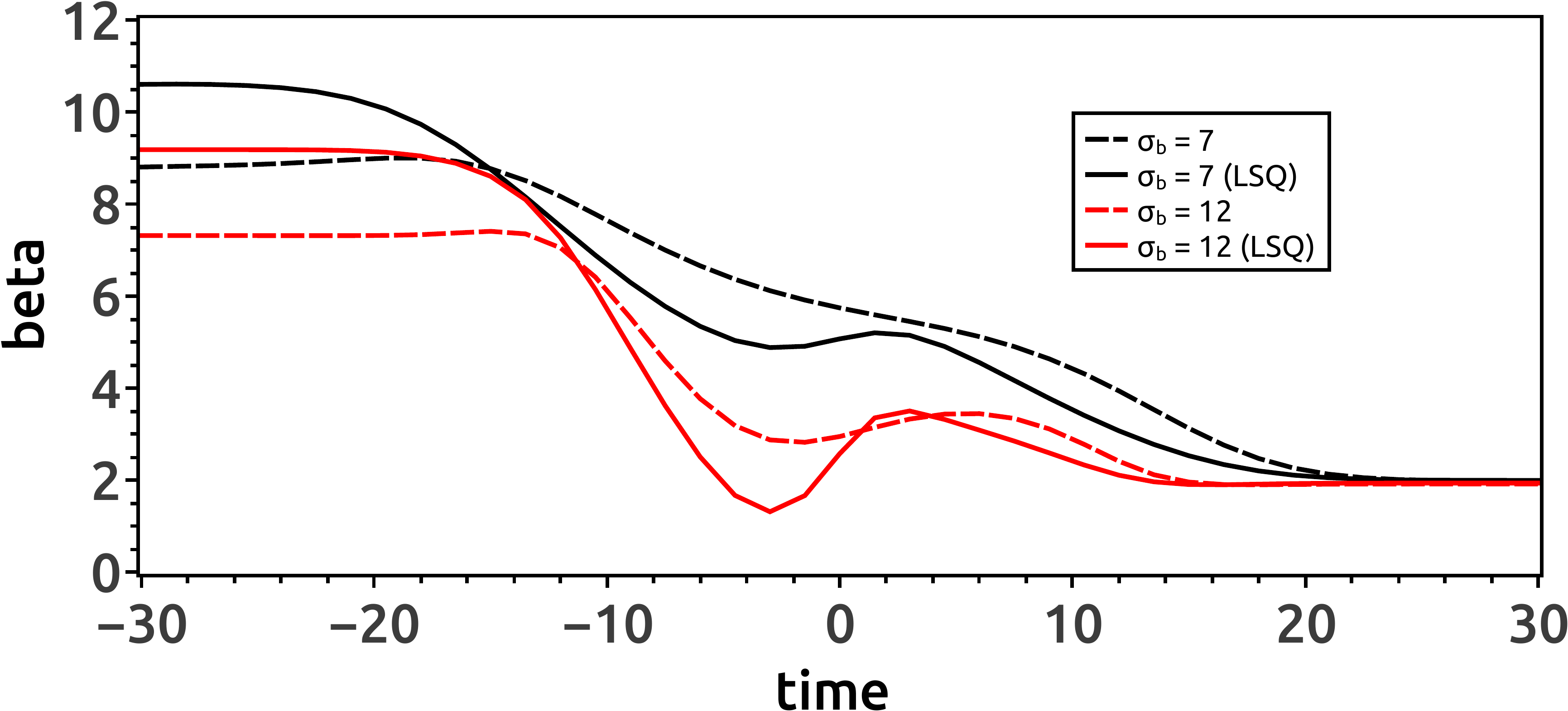}\hfill\includegraphics[width=0.47\linewidth]{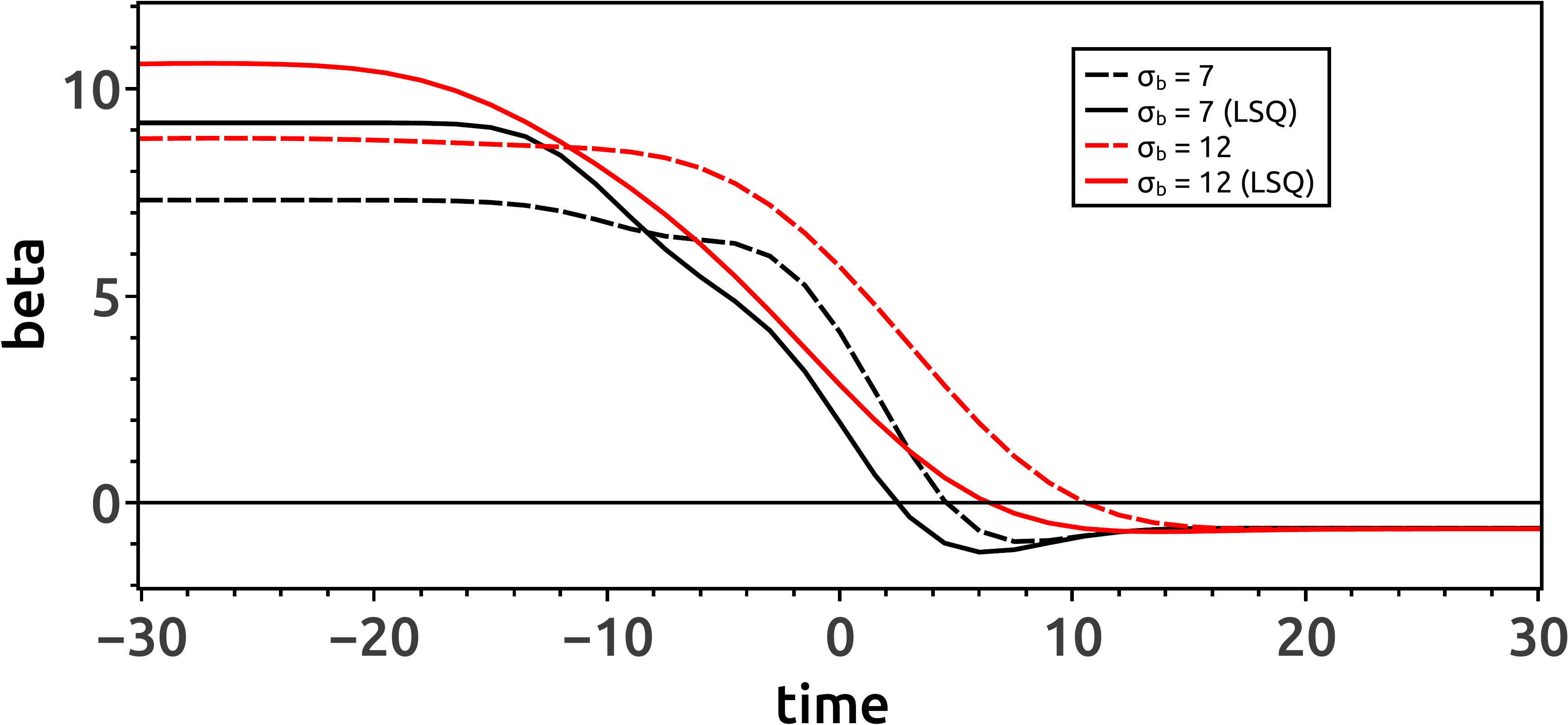}
	\caption{(Colour online) Effective inverse temperatures $\beta=1/T$ determined from the slope of the probe-modified nonequilibrium distribution function at the Fermi level (dashed lines) and by  LSQ interpolation (solid lines) for $E_0=30$ (left) and $E_0=1$ (right).}
	\label{fig:beta}
\end{figure*}

In figures~\ref{fig:pese030} and \ref{fig:pese001}, we plot results for the DOS, PES, and occupation of the single-particle states $n_\text{d}(\omega;t_0)$ for different pump amplitudes and widths. For the large pump amplitude $E_0=30$ (figure~\ref{fig:pese030}), the DOS, which does not depend on temperature, has the same Gaussian profile  far before and far after the pump, but becomes narrowed and sharper near the pump maximum. It also has some small Floquet-like peaks caused by both the pump driving frequency and by the Bloch oscillations of the band energy $\varepsilon(\mathbf{k}-\mathbf{A}(t))$. 

The behaviour for the PES is different. The  PES first displays an equilibrium profile at $t_0=-30$. As we approach the pump maximum at $t_0=0$, the peak becomes narrow and shifts toward zero frequency. At large times, the peak spreads again, but remains in the vicinity of $\omega=0$. 
The difference in the PES spectra before and after the pump is determined by the change in the occupation of the single-particle states. The initial occupation follows the probe-modified Fermi-Dirac distribution. As the pump amplitude increases, the distribution function becomes more flat and exhibits oscillations. After the pump, a step-like shape of the Fermi-Dirac distribution function is almost restored. 
For frequencies in the vicinity of the chemical potential, one can estimate an effective temperature from the slope of the measured distribution function at $\omega=0$ or by using LSQ interpolation (see figure~\ref{fig:beta}).  The initial fitted temperature is close to the actual  initial temperature $\beta=10$ ($T=0.1$); the differences arise from our use of a Fermi-Dirac distribution function instead of a probe-modified distribution function for the fits. During and after the pump, the effective temperature increases and its ``measured'' value is sensitive to the probe-function width $\sigma_{\text{b}}$.

For small amplitudes of the pump, such as $E_0=1$ (see figure~\ref{fig:pese001}), the DOS is weakly modified by the pump while the PES shows a much faster evolution during the pump than it occurs for larger pump amplitudes. Even more strange is the occupation of single-particle states which eventually display population inversion resulting in a negative effective temperature, as seen in figure~\ref{fig:beta}. This population inversion remains after the pump because there are no interactions or other mechanisms for relaxation and thermalization. 

We explain this odd behaviour for small field amplitudes in the following way: The PES is sensitive to the final value of the vector potential because a nonzero value means that the system will remain in a current carrying state (recall that noninteracting metals are perfect conductors). The final value of the pump vector potential is equal to
\begin{equation}
A(+\infty) = -\sqrt{\piup} E_0 \sigma_{\mathrm{p}} \exp\left(-\frac{\omega_{\mathrm{p}}^2\sigma_{\mathrm{p}}^2}{4}\right).
\end{equation}
The induced electric current as a function of the pump vector potential is
\begin{align}
j(t) &= -\ri \int \rd\varepsilon \int \rd\bar{\varepsilon} \rho(\varepsilon,\bar{\varepsilon})  \left[-\varepsilon\sin A_{\text{pump}}(t) + \bar{\varepsilon}\cos A_{\text{pump}}(t)\right] G_{\varepsilon,\bar{\varepsilon}}^{<}(t,t)
=j_0 \sin A_{\text{pump}}(t) 
\end{align}
with
\begin{equation}\label{eq:j0}
j_0 = -\int \rd\varepsilon \dfrac{\re^{-\varepsilon^2}}{\sqrt{\piup}} f(\varepsilon-\mu)\varepsilon = \frac{1}{2}\int \rd\varepsilon \dfrac{\re^{-\varepsilon^2}}{\sqrt{\piup}} \left[-\frac{\rd f(\varepsilon-\mu)}{\rd \varepsilon}\right].
\end{equation}
\begin{figure}[!t]
	\centering
	\includegraphics[width=0.29\linewidth]{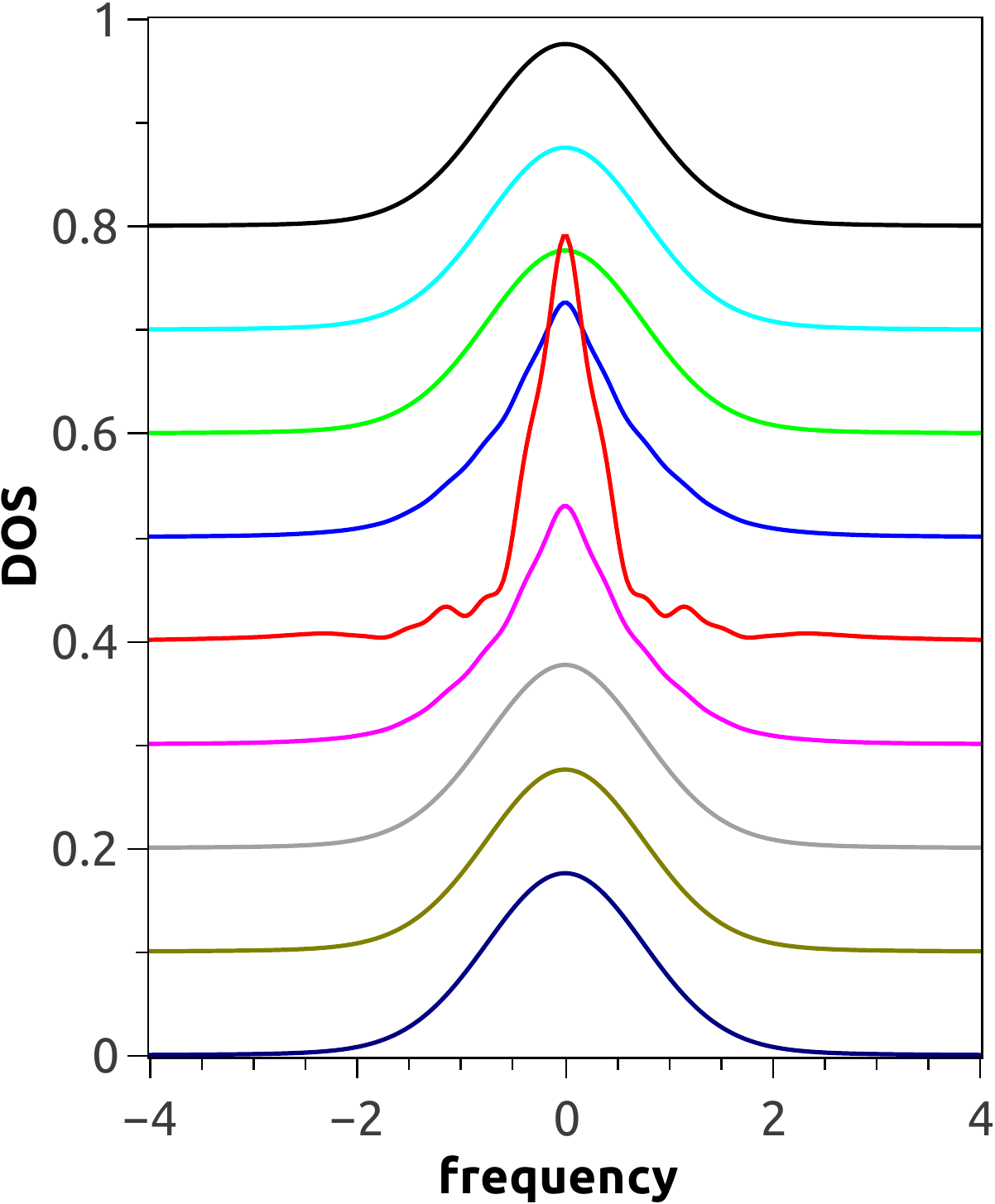}\hfill\includegraphics[width=0.29\linewidth]{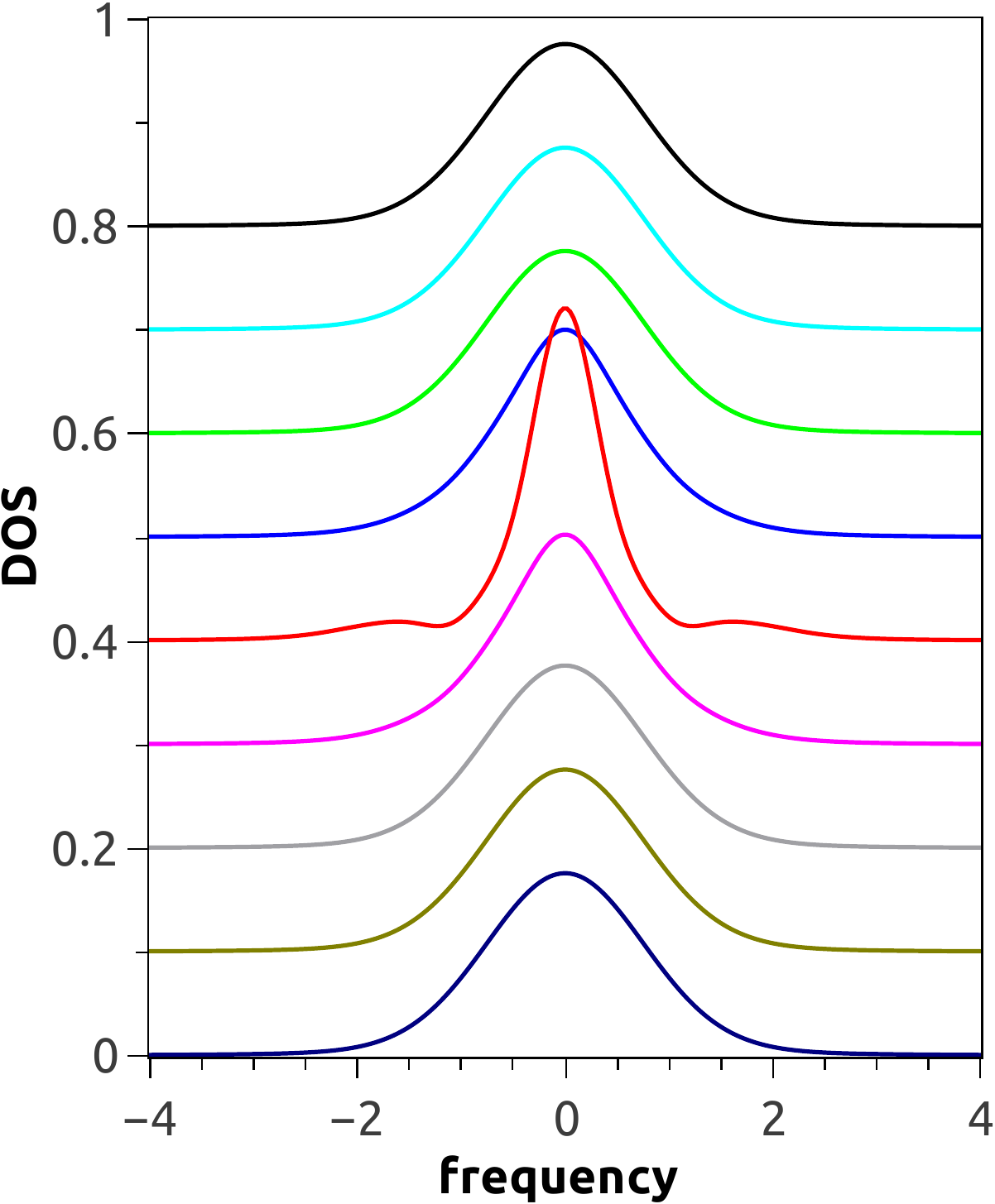}\hfill\includegraphics[width=0.29\linewidth]{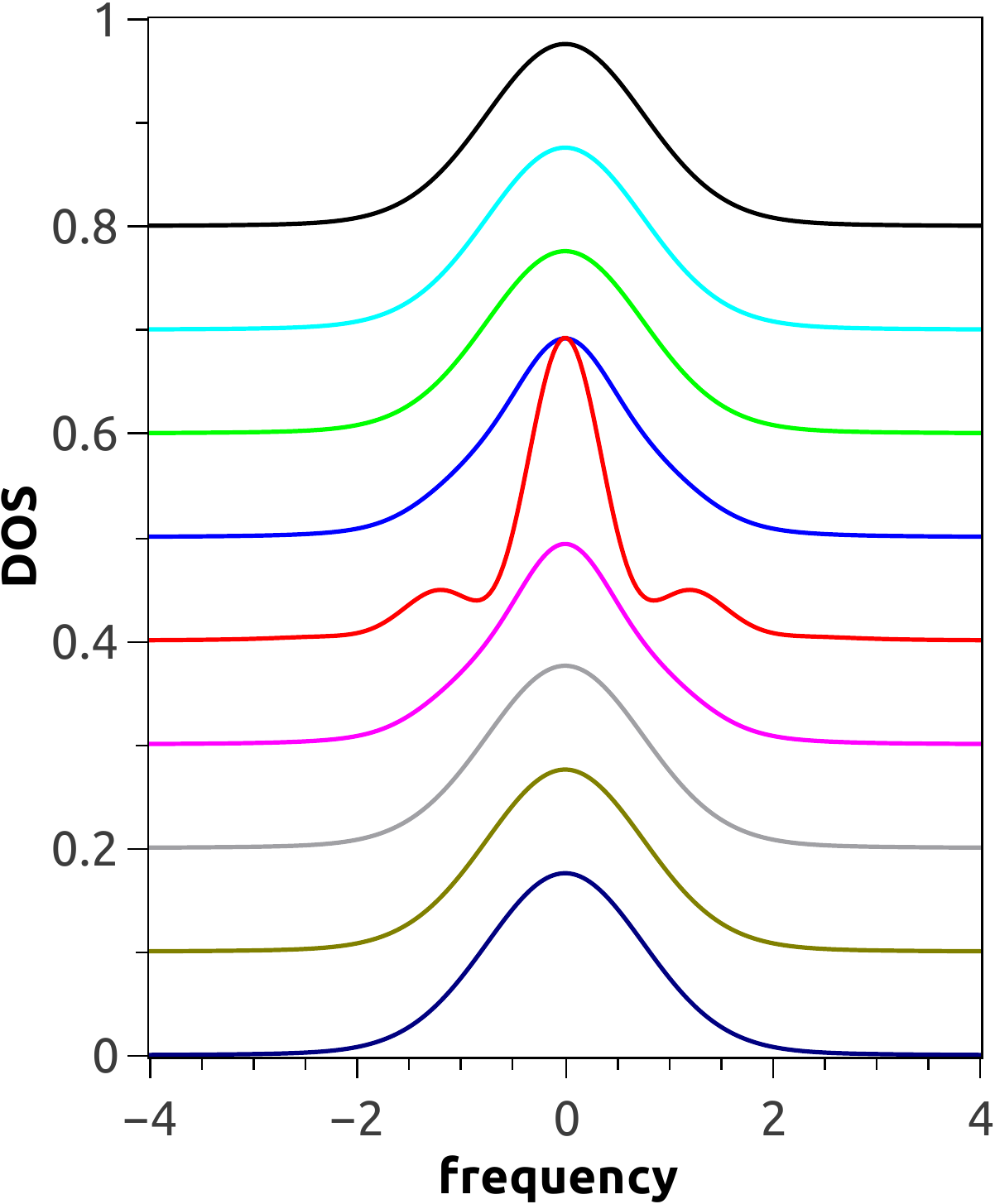} \\ [1em]
	\includegraphics[width=0.29\linewidth]{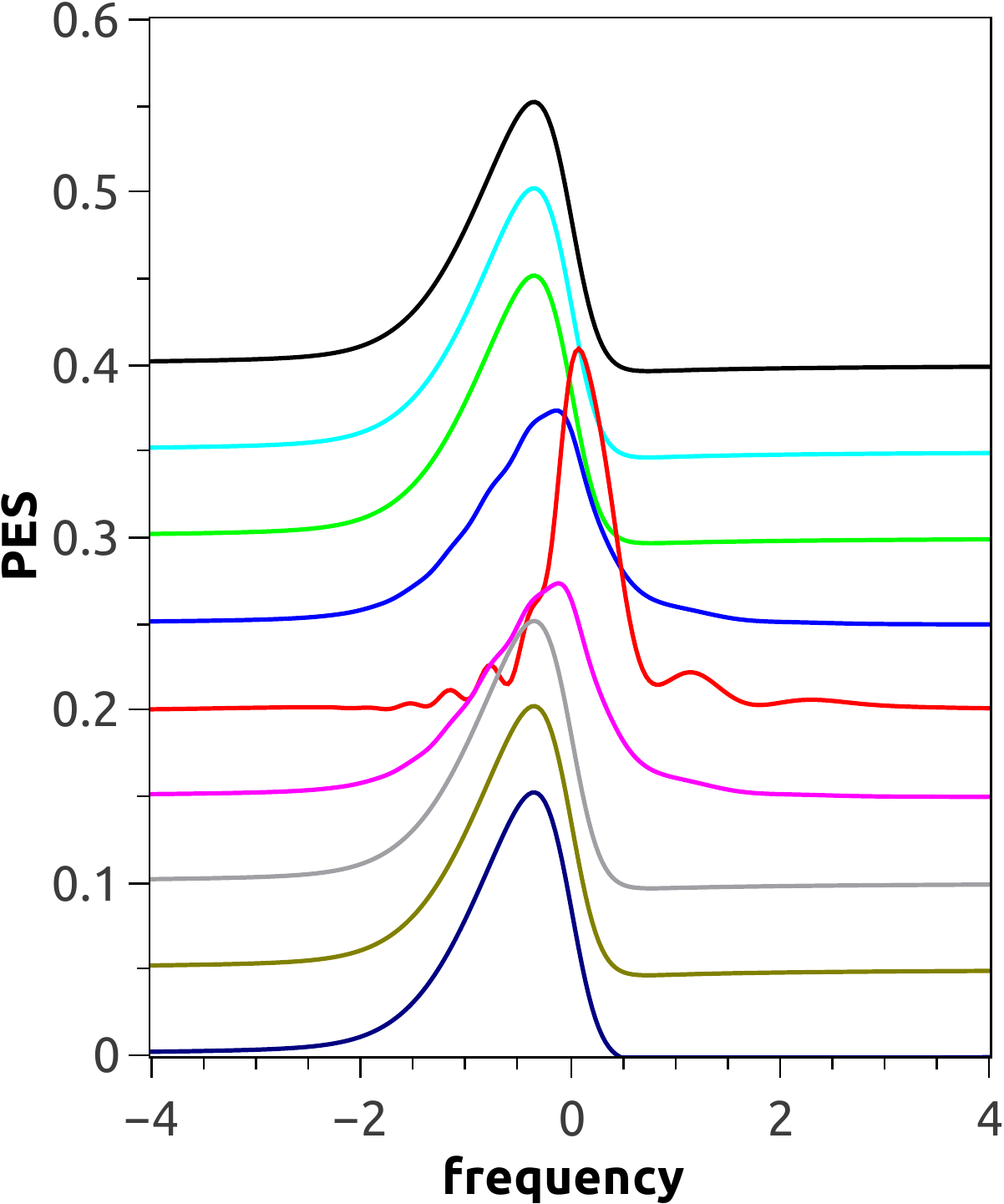}\hfill\includegraphics[width=0.29\linewidth]{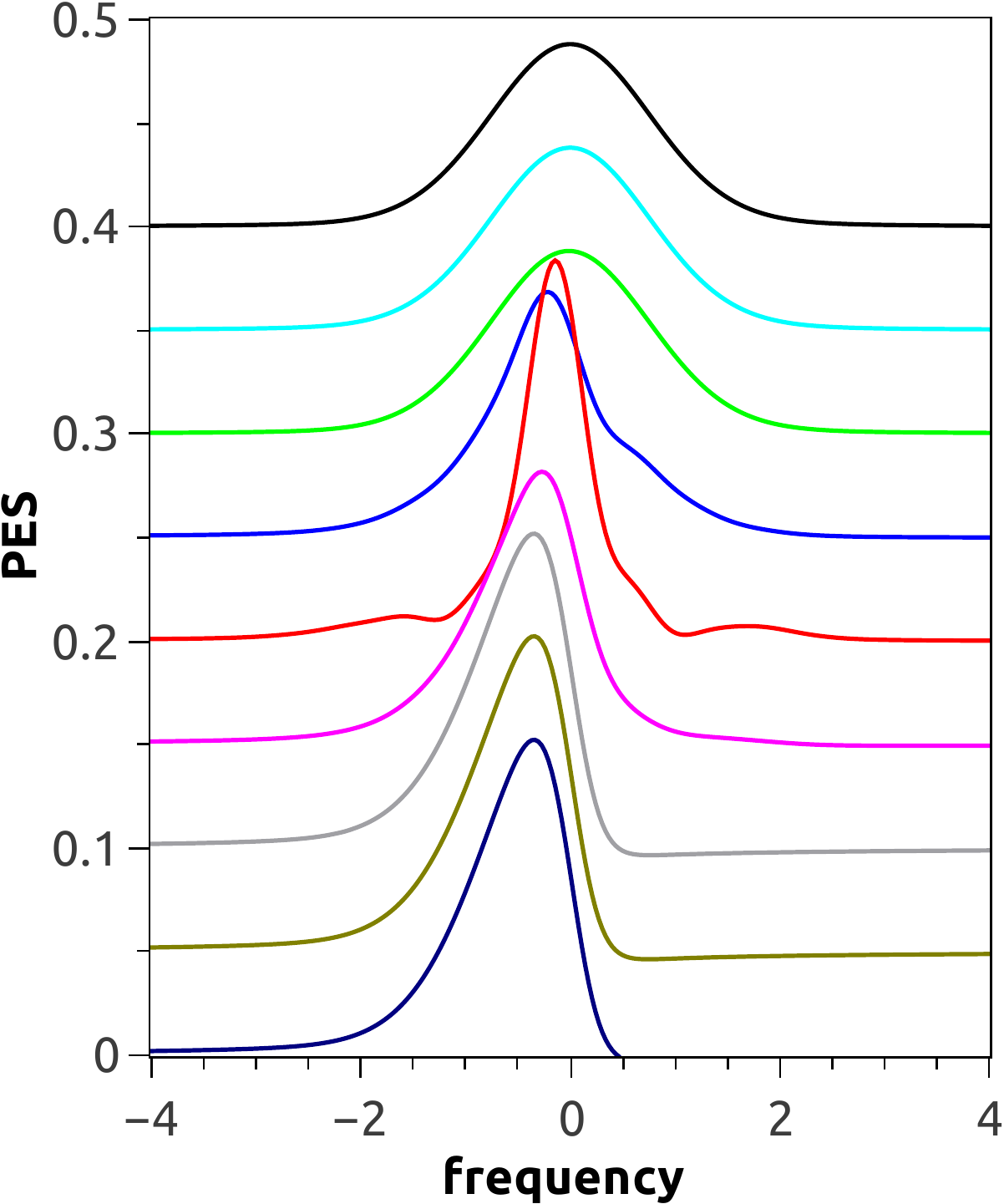}\hfill\includegraphics[width=0.29\linewidth]{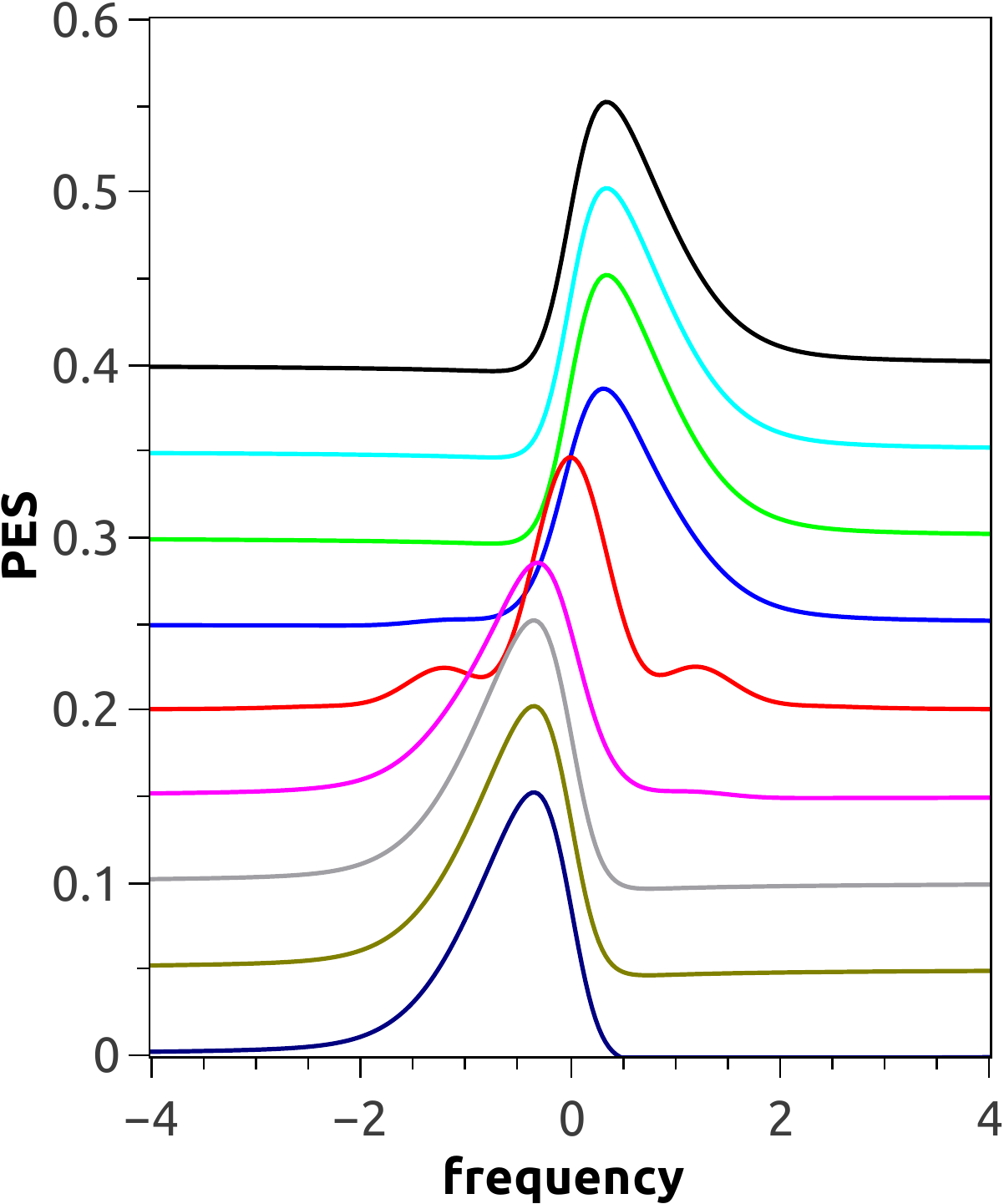} \\ [1em]
	\includegraphics[width=0.33\linewidth]{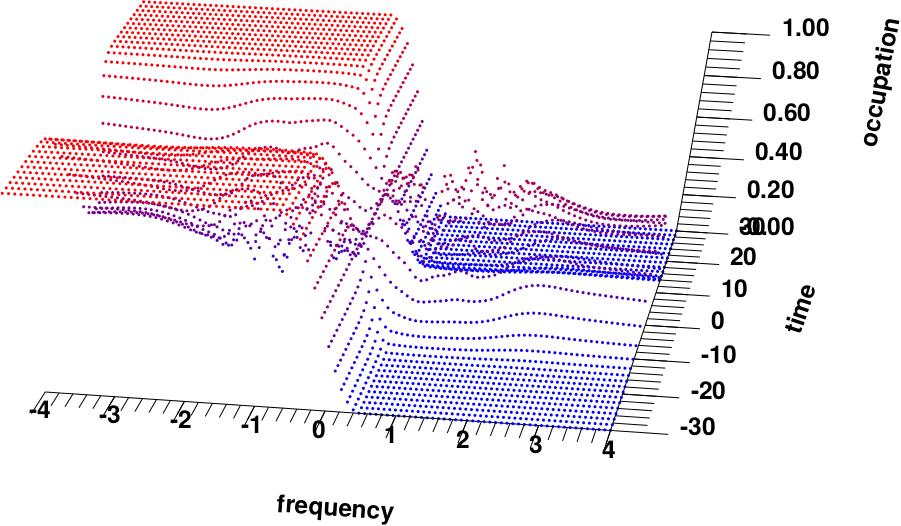}\hfill\includegraphics[width=0.33\linewidth]{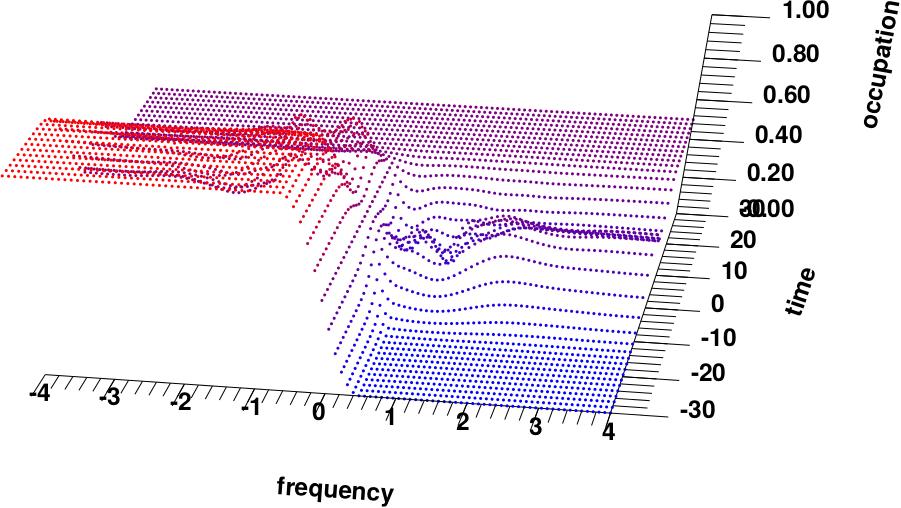}\hfill\includegraphics[width=0.33\linewidth]{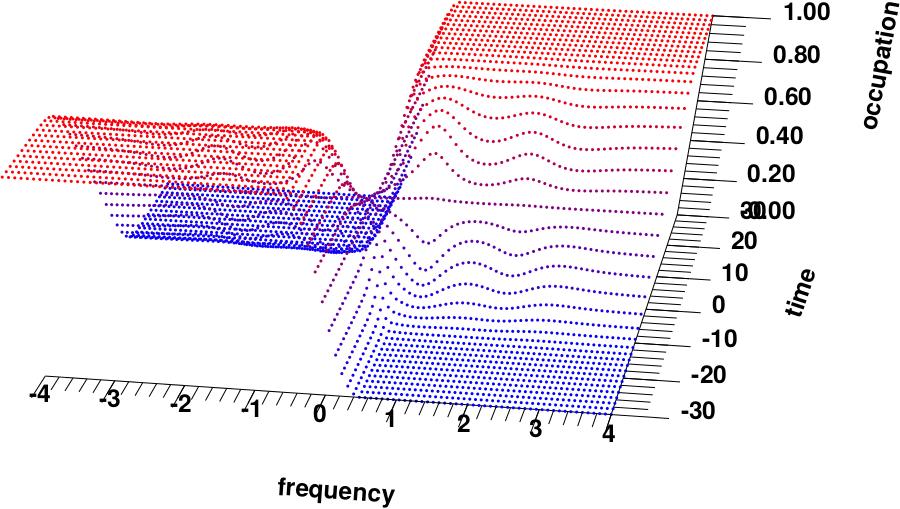} \\ [1em]
	\caption{(Colour online) Waterfall images of the time-resolved DOS and PES data, plotted for different delay times $t_0\in[-30,30]$ and offset for clarity, and occupation of the single particle states $n_\text{d}(\omega)$ for $A(+\infty)=-2\piup$ (left), $-3\piup/2$ (center) and $-\piup$ ($E_0=1.69118088$, right) for $\sigma_{\mathrm{p}}=5$ and $\omega_{\mathrm{p}}=0.5$.}
	\label{fig:pese_pi}
	\bigskip
	\centering
	\includegraphics[width=0.68\linewidth]{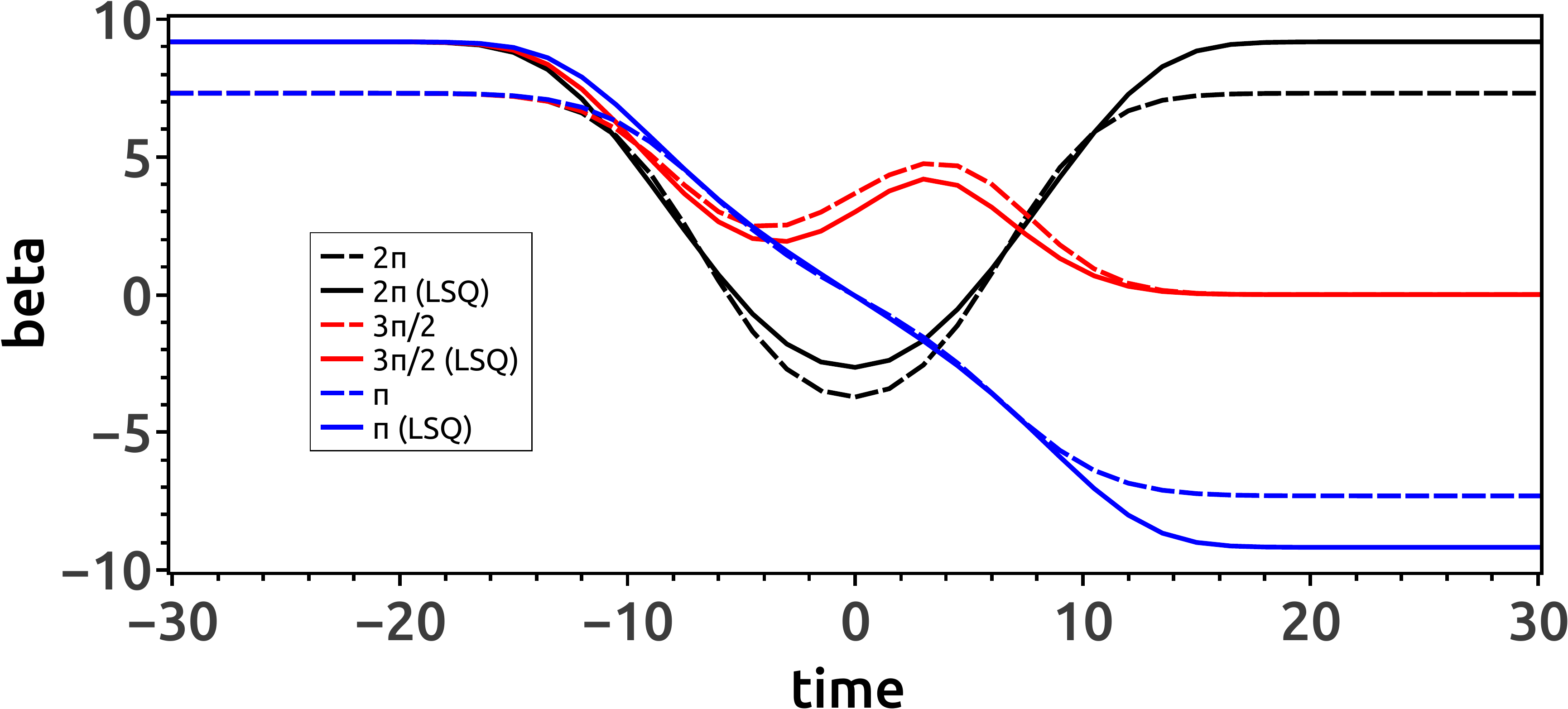}
	\caption{(Color online) Effective inverse temperatures $\beta=1/T$ for $A(+\infty)=-2\piup$, $-3\piup/2$, and $-\piup$.}
	\label{fig:beta_n_pi}
\end{figure}
\par\noindent In figure~\ref{fig:pese_pi}, we present the results for three different pump-field amplitudes which give final vector potential values equal to $A(+\infty)=-2\piup$, $-3\piup/2$, and $-\piup$, respectively.
For $A(+\infty)=2n\piup$, which does not change the final band energy $\varepsilon(\mathbf{k}-\mathbf{A}(+\infty))=\varepsilon(\mathbf{k})$, there is no net current $j(+\infty)=0$ and the final PES (as well as the final occupation of the single-particle states) after the pump is the same as the initial one. Whereas, for $A(+\infty)=(2n+1)\piup$, which changes the sign of the final band energy $\varepsilon(\mathbf{k}-\mathbf{A}(+\infty))=-\varepsilon(\mathbf{k})$ (band flip), we find that the final PES and final occupation of the single-particle states are both inverted even though the net current is still zero $j(+\infty)=0$. The case of $A(+\infty)=\frac{\piup}{2}+n\piup$ makes the band energy antisymmetric (cosine is replaced by sine) so that the net current reaches its maximum allowed value $j(+\infty)=\pm j_0$ and the final PES is half of the DOS with an uniform occupation of the single-particle states $n_\text{d}(\omega)=\frac12$. This appears odd because the PES looks like an infinite-temperature result, while the current is nonzero. But what is happening is we are fully occupying the electrons that move in the direction of the field with no electrons moving opposite. This then saturates the current at its maximum value while making the PES appear to be an infinite-temperature PES because the band degeneracy says for every state moving in the direction of the field that there is a degenerate state moving opposite to the field.
The corresponding changes of the effective inverse temperatures are shown in figure~\ref{fig:beta_n_pi}.
The cases shown in figure~\ref{fig:beta} correspond to both large $E_0=30$ and small $E_0=1$ field amplitudes, with final vector potential values given by $A(+\infty)=-17.74\piup$ and $A(+\infty)=-0.59\piup$, respectively.
The DOS displays a simple behaviour. It has a monotonous change with increasing pump amplitude $E_0$. It has the same equilibrium profiles at long times while it becomes enhanced and narrowed near the pump maximum.

\section{Nonresonant Raman scattering}\label{sec:2particle}

Now, we proceed to examine electronic inelastic light (Raman) scattering, which measures the two-particle excitations. Since a time-varying Hamiltonian does not have well-defined energy eigenstates, we cannot directly apply the Kramers-Heisenberg formula as is often done in the linear-response regime. Instead,  we have to derive the scattering cross-section from scratch using the Nozi\`eres and Abrahams approach~\cite{nozieres:3099} for inelastic light scattering. 

We start by analyzing the evolution of the initial electronic state plus one initial probe photon that has momentum $\mathbf{k}_\text{i}$, polarization $\mathbf{e}_\text{i}$, and frequency $\omega_\text{i} = c |\mathbf{k}_\text{i}|$. Expanding the full evolution operator to include up to the  second-order terms in the probe field, we find
\begin{align}
|\psi(t)\rangle&=U(t,-\infty)|n\rangle\otimes a_{\mathbf{k}_\text{i},\mathbf{e}_\text{i}}^{\dag}|0\rangle  =\mathcal{T}_t\exp{\biggl[-\ri\int^{t}_{-\infty}\rd\tilde{t} \; \mathcal{H}(\tilde{t})\biggr]}\; |n\rangle\otimes a_{\mathbf{k}_\text{i},\mathbf{e}_\text{i}}^{\dag}|0\rangle
\nonumber \\
&\approx\frac{1}{2}\int^{t}_{-\infty}\rd\tilde{t} \; U_{0}(+\infty,\tilde{t})A^{\alpha}_{\text{probe}}(\tilde{t}) \gamma_{\alpha\beta}(\tilde{t})A^{\beta}_{\text{probe}}(\tilde{t})
U_{0}(\tilde{t},-\infty)\; |n\rangle\otimes a_{\mathbf{k}_\text{i},\mathbf{e}_\text{i}}^{\dag}|0\rangle 
\nonumber \\
&+\int^{t}_{-\infty}\rd\tilde{t} \int^{\tilde{t}}_{-\infty}\rd\tilde{t}' \; U_{0}(+\infty,\tilde{t}) j_{\alpha}(\tilde{t})A^{\alpha}_{\text{probe}}(\tilde{t}) U_{0}(\tilde{t},\tilde{t}') j_{\beta}(\tilde{t}')A^{\beta}_{\text{probe}}(\tilde{t}')
U_{0}(\tilde{t},-\infty)\;|n\rangle\otimes a_{\mathbf{k}_\text{i},\mathbf{e}_\text{i}}^{\dag}|0\rangle ,
\label{eq:scattmatr}
\end{align}
where 
\begin{equation}
\gamma_{\alpha\beta}(t)=\sum_{k}\frac{\partial^{2} \varepsilon(\mathbf{k}-\mathbf{A}_{\text{pump}}(t))}{\partial k_{\alpha}\partial k_{\beta}}c_{\mathbf{k}}^{\dag}c_{\mathbf{k}}^{\phantom\dagger}
\end{equation}
is the nonequilibrium generalization of the stress tensor and 
\begin{equation}
j_{\alpha}(t)=\sum_{k}\frac{\partial \varepsilon(\mathbf{k}-\mathbf{A}_{\text{pump}}(t))}{\partial k_{\alpha}}c_{\mathbf{k}}^{\dag}c_{\mathbf{k}}^{\phantom\dagger}
\end{equation}
is the nonequilibrium generalization of the current operator. Note that the time-evolution operator acts on both terms in the tensor product depending on whether it is an electron or photon operator and the operators are in the interaction representation with respect to the photon operator, because the evolution with respect to the electronic and electron-photon coupling terms is included explicitly via the $U_0$ factors (which only include the pump vector potential). In equation~\eqref{eq:scattmatr}, the first term describes nonresonant scattering and the second term describes resonant scattering of the probe photons (within a time envelope)
\begin{equation}
A^{\alpha}_{\text{probe}}(t) = s(t;t_0) \sum_{\mathbf{k}, \mathbf{e}} \left(\frac{2\piup}{\omega_{\mathbf{k}}}\right)^{1/2} e_{\alpha}\left(\re^{\ri\omega_{\mathbf{k}} t} a_{\mathbf{k},\mathbf{e}}^{\dag}+\re^{-\ri\omega_{\mathbf{k}} t} a_{\mathbf{k}, \mathbf{e}}^{\phantom{\dagger}}\right).
\end{equation}
The scattering cross-section is defined by the probability to find a scattered probe photon with momentum~$\mathbf{k}_\text{f}$, polarization $\mathbf{e}_\text{f}$,  and frequency $\omega_\text{f} = c |\mathbf{k}_\text{f}|$ in the final state. It becomes
\begin{equation*}
R=\sum\limits_{\psi}\frac{\re^{-\beta E_\psi}}{\mathcal{Z}} 
\langle \psi(t\to+\infty)|a_{\mathbf{k}_\text{f},\mathbf{e}_\text{f}}^{\dag}a_{\mathbf{k}_\text{f},\mathbf{e}_\text{f}}^{\phantom\dagger}|\psi(t\to+\infty)\rangle,
\end{equation*}
where $\mathcal{H}(t\to -\infty)|\psi(t\to-\infty)\rangle=E_\psi|\psi(t\to-\infty)\rangle$.

After tracing over the photon operators, we find that the nonresonant contribution to the electronic Raman scattering (with frequency loss $\Omega = \omega_\text{i} - \omega_\text{f}$) becomes
\begin{equation}\label{eq:Raman}
R_N(\Omega;t_0)= \ri \int \rd t \int \rd t'\; s^2(t;t_0) s^2(t';t_0) \re^{\ri\Omega(t-t')} R_N(t,t').
\end{equation}
This result arises from the greater Green's function 
\begin{equation}\label{eq:Rgenexp}
R_N(t,t') = R^{>}(t,t') = R^{-+}(t,t') = -\ri\left\langle \tilde{\gamma}(t) \tilde{\gamma}(t')\right\rangle,
\end{equation}
where ($+,-$) denote the upper and lower branches of the Keldysh contour and
\begin{equation}
\tilde{\gamma}(t) = e_{\text{i}\alpha} \gamma_{\alpha\beta}(t) e_{\text{f}\beta}
\end{equation}
is the contraction of the stress tensor with the polarization vectors of the initial (i) and scattered (f) photons.

\subsection{Ratio of Stokes and anti-Stokes peaks and the probe pulse shape}

Let us first consider the influence of the probe pulse shape on the ratio of the Stokes and anti-Stokes peaks~\cite{hayes2012scattering} for the ``equilibrium'' linear-response case. This ratio is often used for the estimation of local temperatures~\cite{kip:90} and can be applied in the pump-probe experiments too~\cite{yang:40876}, but the conventional derivation holds only for the equilibrium case with continuous probes.
In equilibrium, the greater Green's function depends only  on the time difference of its arguments
\begin{equation}
R^{>}(t,t') = R_{\textrm{eq}}^{>}(t-t'),
\end{equation}
hence, the  Fourier transform becomes
\begin{equation}
R_N^{\textrm{eq}}(\Omega) = \ri \int_{-\infty}^{+\infty} \rd (t-t') \; \re^{\ri\Omega (t-t')} R_{\textrm{eq}}^{>}(t-t').
\end{equation}
From the spectral properties of the greater Green's function it follows that the ratio of amplitudes of the Stokes and anti-Stokes lines for the nonresonant Raman scattering of a monochromatic beam is equal to
\begin{equation}\label{S_aS}
\frac{R_N^{\textrm{eq}}(\Omega)}{R_N^{\textrm{eq}}(-\Omega)} = \exp(\beta\Omega).
\end{equation}

Now, let us check how this ratio is distorted by finite-width probe pulse envelope functions. From equation \eqref{eq:Raman}, we obtain (after introducing the inverse Fourier transform for $R_N$)
\begin{align}
R_N(\Omega) &= \int_{-\infty}^{+\infty} \rd t \int_{-\infty}^{+\infty} \rd t' R_{\textrm{eq}}^{>}(t-t') s^2(t;t_0) s^2(t';t_0) \re^{\ri\Omega (t-t')}
\nonumber \\
&= \frac{1}{2\piup} \int_{-\infty}^{+\infty} \rd\Omega' R_N^{\textrm{eq}}(\Omega') \int_{-\infty}^{+\infty} \rd t \int_{-\infty}^{+\infty} \rd t' s^2(t) s^2(t') \re^{\ri(\Omega-\Omega') (t-t')},
\end{align}
where the dependence on the probe pulse time $t_0$ vanishes just like it did for PES because the remainder of the integrand is just a function of relative time.
After substituting in the Gaussian form for the probe-pulse envelope function, we find
\begin{align}
R_N(\Omega) &= \frac{1}{2\piup} \int_{-\infty}^{+\infty} \rd\Omega' R_N^{\textrm{eq}}(\Omega') \frac{1}{\sigma_{\mathrm{b}}^4\piup^2} \int_{-\infty}^{+\infty} \rd t \re^{-2t^2/\sigma_{\mathrm{b}}^2+\ri(\Omega-\Omega') t} \int_{-\infty}^{+\infty} \rd t' \re^{-2t'^2/\sigma_{\mathrm{b}}^2-\ri(\Omega-\Omega') t'}
\nonumber \\
&= \frac{1}{2\piup} \int_{-\infty}^{+\infty} \rd\Omega' R_N^{\textrm{eq}}(\Omega') \re^{-\sigma_{\mathrm{b}}^2(\Omega-\Omega')^2/4} \frac{1}{2\piup\sigma_{\mathrm{b}}^2}.
\label{eq:RNenvlp}
\end{align}
Evaluating equation~\eqref{eq:RNenvlp} for negative frequencies gives
\begin{align}
R_N(-\Omega) &= \frac{1}{4\piup^2\sigma_{\mathrm{b}}^2} \int_{-\infty}^{+\infty} \rd\Omega' R_N^{\textrm{eq}}(-\Omega') \re^{-\sigma_{\mathrm{b}}^2(\Omega-\Omega')^2/4}
= \frac{1}{4\piup^2\sigma_{\mathrm{b}}^2} \int_{-\infty}^{+\infty} \rd\Omega' R_N^{\textrm{eq}}(\Omega') \re^{-\beta\Omega'} \re^{-\sigma_{\mathrm{b}}^2(\Omega-\Omega')^2/4}
\nonumber \\
&= \frac{1}{4\piup^2\sigma_{\mathrm{b}}^2} \re^{-\beta\Omega+\beta^2/\sigma_{\mathrm{b}}^2} \int_{-\infty}^{+\infty} \rd\Omega' R_N^{\textrm{eq}}(\Omega') \re^{-\sigma_{\mathrm{b}}^2(\Omega-\Omega'-2\beta/\sigma_{\mathrm{b}}^2)^2/4}
\nonumber \\
&= \re^{-\beta\left(\Omega-\frac{\beta}{\sigma_{\mathrm{b}}^2}\right)} R_N\left(\Omega-\frac{2\beta}{\sigma_{\mathrm{b}}^2}\right),
\label{eq:RNaSt}
\end{align}
after using the equilibrium relation for the ratio inside the integrand.
We now introduce the shifted frequency $\Tilde\Omega = \Omega - \frac{\beta}{\sigma_{\mathrm{b}}^2}$ and finally conclude that the Stokes-anti-Stokes ratio changes to
\begin{equation}
\frac{R_N\left(\Tilde\Omega-\frac{\beta}{\sigma_{\mathrm{b}}^2}\right)}{R_N\left(-\Tilde\Omega-\frac{\beta}{\sigma_{\mathrm{b}}^2}\right)} =  \exp(\beta\Tilde\Omega),
\end{equation}
when the probe-pulse has a finite width. Note that this shift vanishes at high temperatures, but can be significant at a low temperature.

\subsection{Nonresonant Raman scattering off noninteracting electrons}

We now examine the nonequilibrium case for nonresonant electronic Raman scattering.
For noninteracting fermions, we have only the bare bubble contribution in equation~\eqref{eq:Rgenexp}, which becomes
\begin{align}
R^{>}(t,t') &= \frac{\ri}{N} \sum_{\mathbf{k}} \gamma(\mathbf{k}-\mathbf{A}_{\textrm{pump}}(t)) \gamma(\mathbf{k}-\mathbf{A}_{\textrm{pump}}(t')) G_{\mathbf{k}}^{>}(t,t')G_{\mathbf{k}}^{<}(t',t) 
\nonumber\\
&= \frac{\ri}{N} \sum_{\mathbf{k}} \gamma(\mathbf{k}-\mathbf{A}_{\textrm{pump}}(t)) \gamma(\mathbf{k}-\mathbf{A}_{\textrm{pump}}(t')) G_{\mathbf{k}}^{-+}(t,t')G_{\mathbf{k}}^{+-}(t',t),
\label{R_gt}
\end{align}
where
\begin{equation}
\gamma(\mathbf{k}-\mathbf{A}_{\textrm{pump}}(t)) = \sum_{\alpha,\beta} e_{\text{i}\alpha} \frac{\partial^2\varepsilon(\mathbf{k}-\mathbf{A}_{\textrm{pump}}(t))}{\partial k_{\alpha}\partial k_{\beta}} e_{\text{f}\beta}
\end{equation} 
is the contraction of the stress tensor amplitude with the polarization vector components of the incident (scattered) photons. 

For an $A_{1\mathrm{g}}$ symmetry, with $\mathbf{e}_\text{i}=\mathbf{e}_\text{f}=(1,1,1,\ldots)$, we have $\gamma(\mathbf{k}-\mathbf{A}_{\textrm{pump}}(t))=-\varepsilon(\mathbf{k}-\mathbf{A}_{\textrm{pump}}(t))$, which gives
\begin{align}
R^{>}_{A_{1\mathrm{g}}}(t,t') &= \ri \int \rd\varepsilon \int \rd \bar{\varepsilon}\, \rho(\varepsilon,\bar{\varepsilon}) \left[\varepsilon \cos A_{\text{pump}}(t) + \bar{\varepsilon} \sin A_{\text{pump}}(t)\right] \left[\varepsilon \cos A_{\text{pump}}(t') + \bar{\varepsilon} \sin A_{\text{pump}}(t')\right]
\nonumber \\
&\times G_{\varepsilon,\bar{\varepsilon}}^{-+}(t,t')G_{\varepsilon,\bar{\varepsilon}}^{+-}(t',t).
\label{eq:RneqA1g}
\end{align}
For a $B_{1\mathrm{g}}$ symmetry, with $\mathbf{e}_\text{i}=(1,1,1,1,\ldots)$ and $\mathbf{e}_\text{f}=(1,-1,1,-1,\ldots)$ (and for nearest-neighbour hopping only), we have $\gamma(\mathbf{k}-\mathbf{A}_{\textrm{pump}}(t)) = -\frac{t^*}{\sqrt{D}} \sum_{\alpha=1}^{D} (-1)^{\alpha} \cos [k_{\alpha}-A_{\textrm{pump}}(t)]$ and the nonzero contributions in the $D\to\infty$ limit become
\begin{align}
&\gamma(\mathbf{k}-\mathbf{A}_{\textrm{pump}}(t)) \gamma(\mathbf{k}-\mathbf{A}_{\textrm{pump}}(t')) \to \frac{{t^*}^2}{D} \sum_{\alpha=1}^{D} \cos [k_{\alpha}-A_{\textrm{pump}}(t)] \cos [k_{\alpha}-A_{\textrm{pump}}(t')]
\nonumber \\
&\qquad\qquad= \frac{{t^*}^2}{2}\left\{\cos[A_{\text{pump}}(t)-A_{\text{pump}}(t')] + \frac1D \sum_{\alpha=1}^{D} \cos[2k_{\alpha}-A_{\text{pump}}(t)-A_{\text{pump}}(t')]\right\}
\nonumber \\
&\qquad\qquad\to \frac{{t^*}^2}{2} \cos[A_{\text{pump}}(t)-A_{\text{pump}}(t')],
\end{align}
which we can use to rewrite \eqref{R_gt} as
\begin{equation}
R^{>}_{B_{1\mathrm{g}}}(t,t') = \ri\,\frac{{t^{*}}^2}{2} \cos[A_{\text{pump}}(t)-A_{\text{pump}}(t')] \int \rd\varepsilon \int \rd \bar{\varepsilon}\,\rho(\varepsilon,\bar{\varepsilon}) G_{\varepsilon,\bar{\varepsilon}}^{-+}(t,t')G_{\varepsilon,\bar{\varepsilon}}^{+-}(t',t).
\label{eq:Rneq}
\end{equation}

Remarkably, for noninteracting electrons, the product of the lesser and greater Green's functions \eqref{eq:Gc} is equal to just the products of Fermi factors
\begin{equation}
G_{\varepsilon,\bar{\varepsilon}}^{-+}(t,t')G_{\varepsilon,\bar{\varepsilon}}^{+-}(t',t) = f(\varepsilon-\mu) [1-f(\varepsilon-\mu)]
\end{equation}
and does not depend on the time variables. As a result, an equilibrium Raman cross-section ($A_{\text{pump}}(t)=0$) for a monochromatic beam is proportional to a $\delta$-function, $R^{N}(\Omega) \sim \delta(\Omega)$ (no inelastic light scattering), whereas in a pump/probe experiment, we find 
\begin{align}
R^{N}_{A_{1\mathrm{g}}}(\Omega) &= \int \rd\varepsilon \int \rd \bar{\varepsilon}\, \rho(\varepsilon,\bar{\varepsilon}) f(\varepsilon-\mu) [1-f(\varepsilon-\mu)] \int \rd t \int \rd t'\; s^2(t) s^2(t') \re^{\ri\Omega(t-t')}
\nonumber \\
&\times \left[\varepsilon \cos A_{\text{pump}}(t) + \bar{\varepsilon} \sin A_{\text{pump}}(t)\right] \left[\varepsilon \cos A_{\text{pump}}(t') + \bar{\varepsilon} \sin A_{\text{pump}}(t')\right] 
\nonumber \\
&= \int \rd\varepsilon \frac{\re^{-\varepsilon^2}}{\sqrt{\piup}} f(\varepsilon-\mu) [1-f(\varepsilon-\mu)] \varepsilon^2 \left|\int \rd t \, s^2(t) \re^{\ri\Omega t} \cos A_{\text{pump}}(t)\right|^2
\nonumber \\
&+ \frac{1}{2} T N(\mu) \left|\int \rd t \, s^2(t) \re^{\ri\Omega t} \sin A_{\text{pump}}(t)\right|^2
\label{eq:RneqA1gfin}
\end{align}
and
\begin{align}
R^{N}_{B_{1\mathrm{g}}}(\Omega) &= T N(\mu) \cdot\frac{{t^*}^2}{2}  
\int \rd t \int \rd t'\; s^2(t) s^2(t') \re^{\ri\Omega(t-t')} \cos\left[A_{\text{pump}}(t)-A_{\text{pump}}(t')\right]
\nonumber \\
&= T N(\mu) \cdot\frac{{t^*}^2}{2} \left[\left|\int \rd t \, s^2(t) \re^{\ri\Omega t} \cos A_{\text{pump}}(t)\right|^2 + \left|\int \rd t \, s^2(t) \re^{\ri\Omega t} \sin A_{\text{pump}}(t)\right|^2 \right].
\end{align}
Here, the prefactor $N(\mu)$ does not depend on the pump and only gives the occupation of the Fermi level at $T=0$
\begin{equation}\label{eq:Nmu}
N(\mu) = \beta \int \rd\varepsilon \;\dfrac{\re^{-\varepsilon^2}}{\sqrt{\piup}} f(\varepsilon-\mu) \left[1-f(\varepsilon-\mu)\right] = \int \rd\varepsilon \;\dfrac{\re^{-\varepsilon^2}}{\sqrt{\piup}} \left[-\frac{\rd f(\varepsilon-\mu)}{\rd\varepsilon}\right]=2j_0.
\end{equation}
In equilibrium, as well as far before or far after a pump (where $A_{\text{pump}}(t)=A_{\textrm{eq}}$), we obtain Gaussian profiles for both symmetry channels:
\begin{align}
R^{N}_{A_{1\mathrm{g}}}(\Omega) &= \left\{\int \rd\varepsilon \frac{\re^{-\varepsilon^2}}{\sqrt{\piup}} f(\varepsilon-\mu) [1-f(\varepsilon-\mu)] \varepsilon^2 \cos^2 A_{\textrm{eq}} + \frac{1}{2} T N(\mu) \sin^2 A_{\textrm{eq}}\right\}
\frac{2}{\sigma_{\text{b}}^2\piup} \re^{-\sigma_{\text{b}}^2\Omega^2/4}
\nonumber \\
&\approx \frac{1}{2} T N(\mu)
\frac{2}{\sigma_{\text{b}}^2\piup} \re^{-\sigma_{\text{b}}^2\Omega^2/4} \sin^2 A_{\textrm{eq}} \quad \text{for }\quad T\to0
\label{eq:RneqA1geq}
\end{align}
and
\begin{align}
R^{N}_{B_{1\mathrm{g}}}(\Omega) &= T N(\mu) \cdot\frac{{t^*}^2}{2} 
\frac{2}{\sigma_{\text{b}}^2\piup} \re^{-\sigma_{\text{b}}^2\Omega^2/4}.
\end{align}
This corresponds to the spreading of the $\delta$-peaks by the finite-width probe pulses. One can see that the final $A_{1\mathrm{g}}$ scattering amplitude does depend on the final value of the vector potential, while the $B_{1\mathrm{g}}$ scattering is independent of the final vector potential value.

\begin{figure*}[!t]
	\centering
	\includegraphics[width=0.49\linewidth]{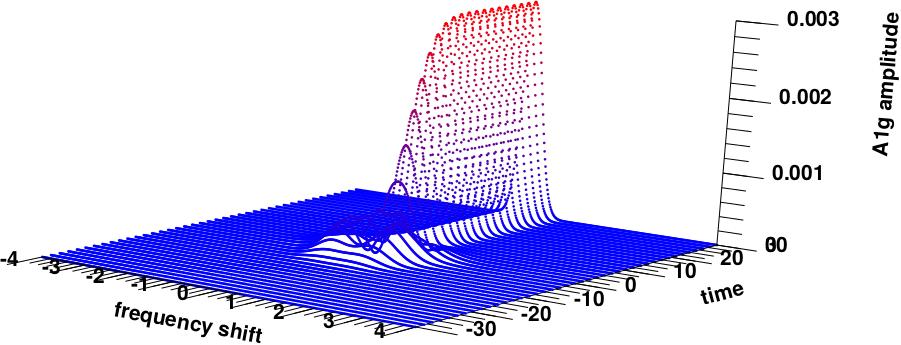}\hfill\includegraphics[width=0.49\linewidth]{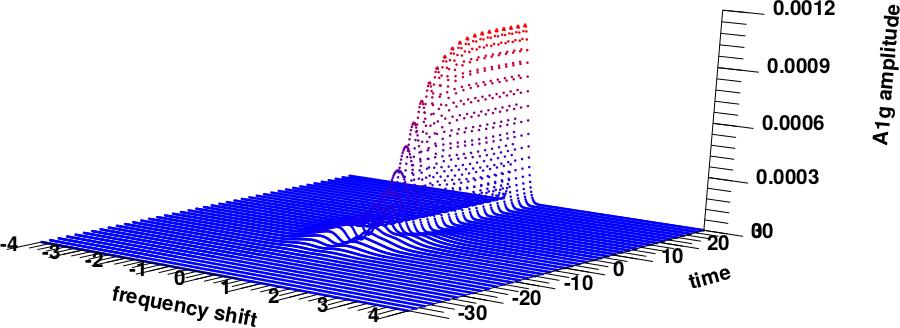} \\ [1em]
	\includegraphics[width=0.49\linewidth]{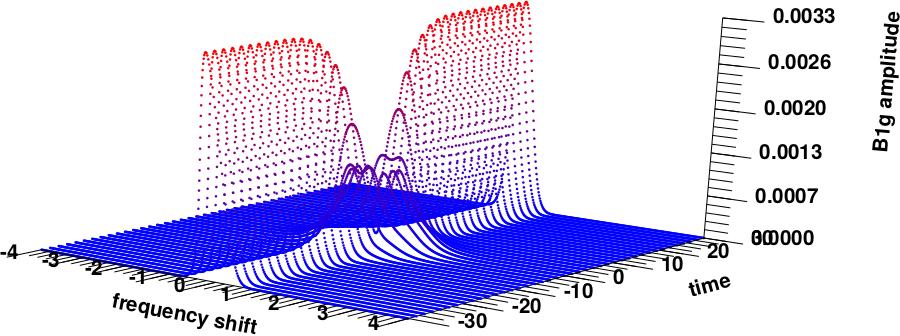}\hfill\includegraphics[width=0.49\linewidth]{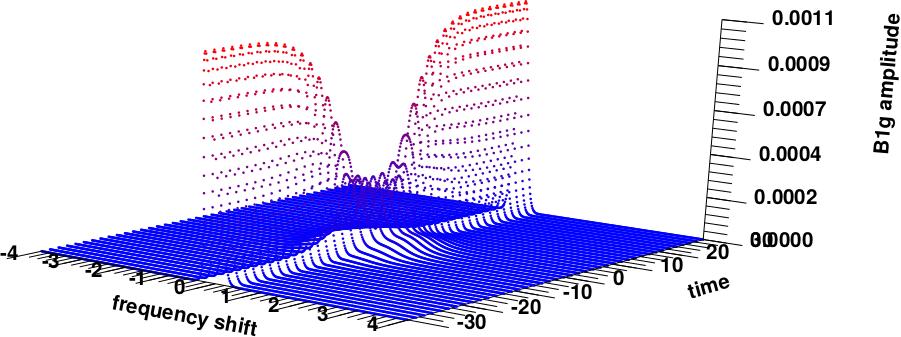}
	\caption{(Colour online) $A_{1\mathrm{g}}$ and $B_{1\mathrm{g}}$ Raman cross sections for $E_0=1$ and $\sigma_{\mathrm{b}}=7$ (left) and $\sigma_{\mathrm{b}}=12$ (right).}
	\label{fig:raman001}
\end{figure*}
\begin{figure*}[!b]
	\centering
	\includegraphics[width=0.49\linewidth]{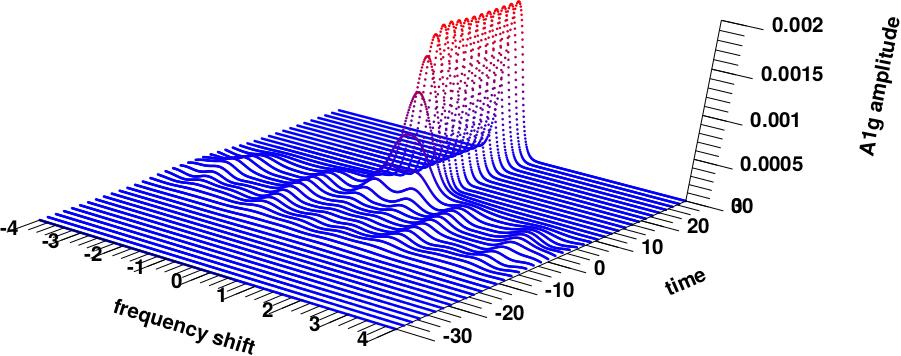}\hfill\includegraphics[width=0.49\linewidth]{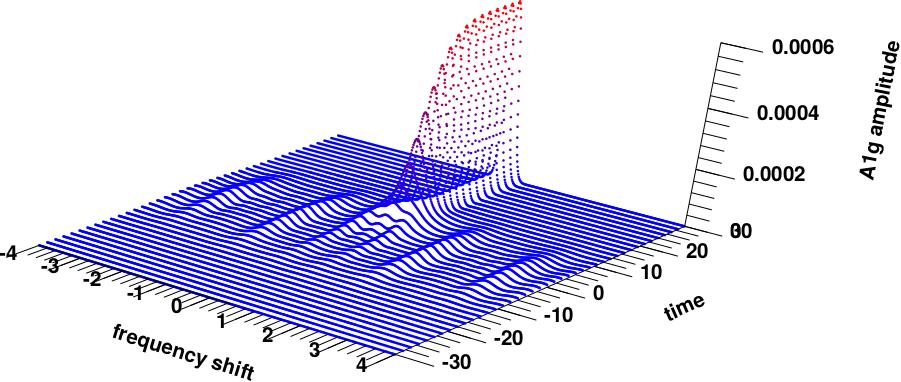} \\ [1em]
	\includegraphics[width=0.49\linewidth]{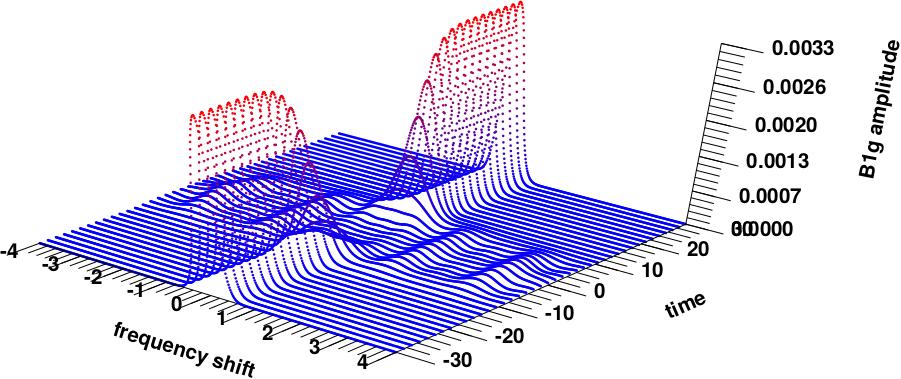}\hfill\includegraphics[width=0.49\linewidth]{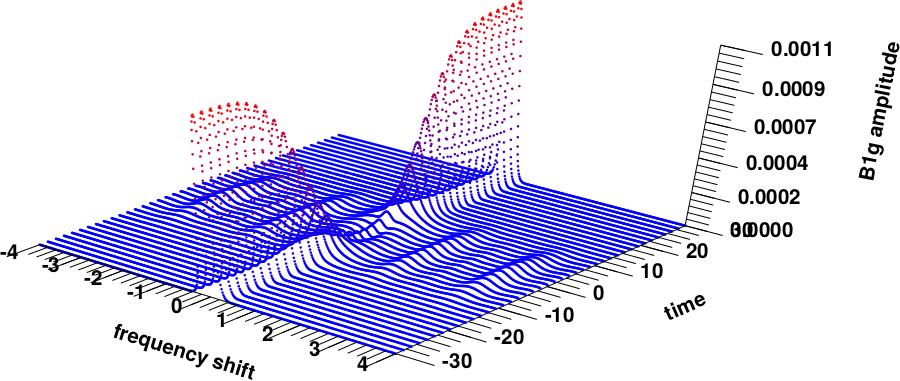}
	\caption{(Colour online) $A_{1\mathrm{g}}$ and $B_{1\mathrm{g}}$ Raman cross sections for $E_0=30$ and $\sigma_{\mathrm{b}}=7$ (left) and $\sigma_{\mathrm{b}}=12$ (right).}
	\label{fig:raman030}
\end{figure*}

If we instead examine the Raman response function
\begin{equation}\label{eq:chi}
\chi_N(\Omega) = R^{N}(\Omega) \left(\re^{\,\beta\Omega}-1\right)
\end{equation}
defined by the retarded Green's function
\begin{equation}\label{eq:Gret}
R^\text{r} (t,t') = -\ri \Theta(t-t')\left\langle\left[\tilde{\gamma}(t),\tilde{\gamma}(t')\right]\right\rangle,
\end{equation} 
we find it is exactly zero
\begin{equation}\label{eq:chi0}
\chi_N(\Omega) \equiv 0.
\end{equation}
This is obviously true for monochromatic beams, where it follows from the identity $\delta(\Omega)(\re^{\,\beta\Omega}-1)=0$, whereas for the finite-width probe pulses, ``the Raman response'' defined by \eqref{eq:chi} is only nonzero due to the spreading of  the $\delta$-peak.

In equilibrium, the Raman cross-section is the product of some prefactor and $\delta(\Omega)$. In  figures~\ref{fig:raman001} and \ref{fig:raman030}, we observe a spreading of the $\delta$-function due to both the finite width of the probe envelope functions and due to the cosine factor (which originated from the stress tensor as modified by pump), which describes the inelastic scattering of light by the temporal variations of the driven stress tensor, due to Bloch oscillations.

Since the Raman response is equal to zero and all of the temperature dependence, as well as the band contributions of the Raman cross-section reside in the factors like $TN(\mu)$, there can be no separation of Stokes and anti-Stokes lines, and the profiles are symmetric. One can see the suppression of the scattering amplitude as well as the appearance of additional sideband oscillations when we are at the pump maximum. These are more prominent for large pump field amplitudes, but the oscillations are more likely an interference effect or \textit{Brillouin} scattering off the time variations of the stress tensor than Floquet sidebands.  It is the oscillations observed in the single-particle DOS and the ``occupations'' at the pump maximum that can be primarily attributed to the Floquet bands~\cite{kalthoff:035138}.

\section{Conclusions}\label{sec:sec:conclusions}

In this article, we have investigated the effect of pulse shapes on pump-probe spectroscopies, on the probe modified nonequilibrium DOS, and PES as well as on electronic Raman scattering. We considered noninteracting fermions on a $D=\infty$ hypercubic lattice with a Gaussian unperturbed DOS, which allows one to obtain some analytic results.

The nonequilibrium DOS $A_\text{d}(\omega;t_0)$ immediately follows the pump pulse. It is completely restored after the pump (to its equilibrium result) and displays both a band narrowing and side-band peaks of Floquet-like sidebands near the pump maximum. For wider probe pulses, more details of the fine structure of the DOS are observed, whereas for narrow probes, the DOS is more smooth; nevertheless, the main peaks are all still visible.

The time evolution of the nonequilibrium PES $P(\omega;t_0)$ is different. After the pump, the PES strongly deviates from the initial one and such a behaviour can be attributed to the nonequilibrium redistribution of the occupations of the single-particle states 
\begin{equation}
n_\text{d}(\omega;t_0) = \dfrac{P(\omega;t_0)}{A_\text{d}(\omega;t_0)}.
\end{equation}
From the slope of the occupation of single-particle states $n_\text{d}(\omega;t_0)$, one can estimate an effective temperature of the single-particle excitations, which is increasing with the pump, and its value depends on the probe width, which modifies the Fermi-Dirac distribution function. For some pump profiles, an inverse occupation of the single-particle states is observed leading to negative effective temperatures.

We have also developed the general approach for obtaining the nonequilibrium Raman cross section and derived an expression for the nonresonant case. We find that even in equilibrium, without a pump, the ratio of Stokes to anti-Stokes peaks is strongly modified by the probe pulse width, and the deviation becomes large at low temperatures and for narrow probe pulses. The Raman response is zero for  noninteracting fermions; nevertheless, the Raman cross section is nonzero and displays an interesting time evolution. For the early and late time values, before and after the pump, the Gaussian central peak, more prominent for the $B_{1\mathrm{g}}$ symmetry, is observed at zero frequency, which is a probe-modified $\delta$-function. With an increasing pump, the central peak is suppressed and splits into a series of peaks whose frequency distribution depends on the pump parameters, the field amplitude $E_0$ and the driving frequency $\omega_{\mathrm{p}}$. We suppose that these peaks are more likely an interference effect or \textit{Brillouin} scattering off the time variations of the stress tensor due to Bloch oscillations.

\section*{Acknowledgements}

It is our pleasure to dedicate this paper to the 80th birthday of Professor I.V.~Stasyuk, a prominent scientist and lecturer, who has made valuable contributions in many fields of quantum statistical physics and solid state theory and taught one of us (A.M.S.) the enjoyment of the Green's functions~\cite{stasyukGFbook}. 

This work was supported by the Department of Energy, Office of Basic Energy Sciences, Division of Materials Sciences and Engineering under Contract Nos. DE-AC02-76SF00515 (Stanford/SIMES) and DE-FG02-08ER46542 (Georgetown).  J.K.F. was also supported by the McDevitt bequest at Georgetown.

\newpage


\newpage

\ukrainianpart

\title{Інтерпретація впливу форми імпульсів у спектроскопії нагнітання-вимір}
\author{А.М. Швайка\refaddr{ICMP}, О.П. Матвєєв\refaddr{ICMP}, Т.П. Деверо\refaddr{Geballe,SIMES}, Дж.К. Фрірікс\refaddr{GU}}
\addresses{
\addr{ICMP} Інститут фізики конденсованих систем НАН України, вул. І. Свєнціцького, 1, 79011 Львів, Україна
\addr{Geballe} Ґебаллівська лабораторія передових матеріалів, Стенфордський університет,\\ Стенфорд, Каліфорнія 94305, США
\addr{SIMES} Стенфордський інститут матеріалів та наук про енергетику (SIMES), Національна прискорювальна лабораторія SLAC, Мелно Парк, Каліфорнія 94025, США
\addr{GU} Фізичний факультет, Джорджтаунський університет,\\ вул. 37 \& О NW, Вашинґтон, округ Колумбія 20057, США
}

\makeukrtitle

\begin{abstract}
	\tolerance=3000%
	Досліджено вплив форми імпульсу в експериментах з імпульсами нагнітання та виміру для випадку найпростішої моделі невзаємодіючих ферміонів на безмежновимірній гіперкубічній ґратці. Отримано, що модифікована імпульсом виміру густина станів слідує часовій еволюції імпульсу нагнітання. Коли імпульс нагнітання досягає максимуму, пік на густині станів вужчає і з'являються додаткові Флоке-подібні бокові зони. Спектри фотоелектронної емісії також зазнають значних змін внаслідок нерівноважного заповнення одночастинкових станів під дією імпульсу нагнітання. Виведено формулу для розрахунку нерівноважного перерізу комбінаційного розсіяння світла та отримано, що нерівноважна складова перерізу розсіяння як для $A_{1\mathrm{g}}$, так і для $B_{1\mathrm{g}}$ симетрій має багатопікову структуру, що може бути пояснено ефектами інтерференції чи бріллюенового розсіяння на часових змінах оператора тензора напружень. Отримано, що і ``виміряне'' заповнення одночастинкових станів, і відношення інтенсивності стоксових до антистоксових піків сильно залежать від ширини пробного імпульсу, що необхідно враховувати при аналізі результатів експериментів.
	\keywords спектроскопія нагнітання-вимір, фотоелектронна емісія, комбінаційне розсіяння світла, нерівноважна функція Ґріна
	
\end{abstract}

\end{document}